\documentclass[journal,twocolumn]{IEEEtran}
\usepackage[numbers,sort&compress]{natbib}
\usepackage{epsfig}
\usepackage{graphicx}
\usepackage{amsmath}
\usepackage{amssymb,subfigure,cases}
\usepackage{setspace}
\usepackage{booktabs} 
\usepackage{color,hyperref}
\usepackage{algorithm}
\usepackage{algorithmicx}
\usepackage{algpseudocode}
\usepackage{multirow}
\usepackage{makecell}
\newtheorem{Definition}{Definition} 
\newtheorem{Lemma}{Lemma}
\newtheorem{theorem}{Theorem}

\usepackage{url}
\usepackage{bm}

\ifCLASSINFOpdf

\else

\fi

\definecolor{darkblue}{rgb}{0.0,0.0,1.0}
\hypersetup{colorlinks,breaklinks,
            linkcolor=darkblue,urlcolor=darkblue,
            anchorcolor=darkblue,citecolor=darkblue}

\begin{document}

\title{Feature Weighted Non-negative Matrix Factorization}

\author{Mulin Chen, Maoguo Gong, and Xuelong Li, \textit{Fellow, IEEE}  % <-this % stops a space
\thanks{
M. Chen is with the Academy of Advanced Interdisciplinary Research, Xidian University, Xi'an 710071, and the Center for OPTical Imagery Analysis and Learning (OPTIMAL), Northwestern Polytechnical University, Xi'an 710072, Shaanxi China. E-mail: chemulin001@gmail.com.

M. Gong is with the Key Laboratory of Intelligent Perception and Image Understanding of Ministry of Education, International Research Center for
Intelligent Perception and Computation, Xidian University, Xi'an 710071, Shaanxi, China. E-mail: gong@ieee.org.

X, Li is with the Center for OPTical Imagery Analysis and Learning (OPTIMAL), Northwestern Polytechnical University, Xi'an 710072, China. E-mail: xuelong$\_$li@nwpu.edu.cn.
}

% <-this % stops a space
%\thanks{Manuscript received April 19, 2005; revised January 11, 2007.}}
}

\markboth{{IEEE} Transactions on XXX}%
{Shell \MakeLowercase{\textit{et al.}}: Bare Demo of IEEEtran.cls for Journals}

\maketitle

\begin{abstract}
Non-negative Matrix Factorization (NMF) is one of the most popular techniques for data representation and clustering, and has been widely used in machine learning and data analysis. NMF concentrates the features of each sample into a vector, and approximates it by the linear combination of basis vectors, such that the low-dimensional representations are achieved. However, in real-world applications, the features are usually with different importances. To exploit the discriminative features, some methods project the samples into the subspace with a transformation matrix, which disturbs the original feature attributes and neglects the diversity of samples. To alleviate the above problems, we propose the Feature weighted Non-negative Matrix Factorization (FNMF) in this paper. The salient properties of FNMF can be summarized as threefold: 1) it learns the weights of features adaptively according to their importances; 2) it utilizes multiple feature weighting components to preserve the diversity; 3) it can be solved efficiently with the suggested optimization algorithm. Performance on synthetic and real-world datasets demonstrate that the proposed method obtains the state-of-the-art performance.

\end{abstract}


\markboth{IEEE TRANSACTIONS ON Cybernetics}%
{}

\definecolor{limegreen}{rgb}{0.2, 0.8, 0.2}
\definecolor{forestgreen}{rgb}{0.13, 0.55, 0.13}
\definecolor{greenhtml}{rgb}{0.0, 0.5, 0.0}

\section{Introduction}
\label{Introduction}
\IEEEPARstart{D}ATA representation is a fundamental task in the field of machine learning and data mining. Generally speaking, it aims to characterize the original data with informative representations, which are suitable for the further processes, such as clustering and classification. Over the past decades, numerous data representation techniques have been proposed, including Principal Component Analysis (PCA)~\cite{pca}, Locality Preserving Projections (LPP)~\cite{lpp}, Locally Linear Embedding (LLE)~\cite{lle}, Non-negative Matrix Factorization (NMF)~\cite{nmf}, deep representation~\cite{deep}, etc. Among them, NMF has attracted considerable attentions due to its advantages on interpretability, and shown promising performance in real-world applications, such as document clustering~\cite{docu, cf}, biological sequence analysis~\cite{gene} and hyperspectral imagery~\cite{robust1,zhangzihan}.

\begin{figure}
\begin{center}
\subfigure[Class 1]{
\includegraphics[width=0.46\textwidth]{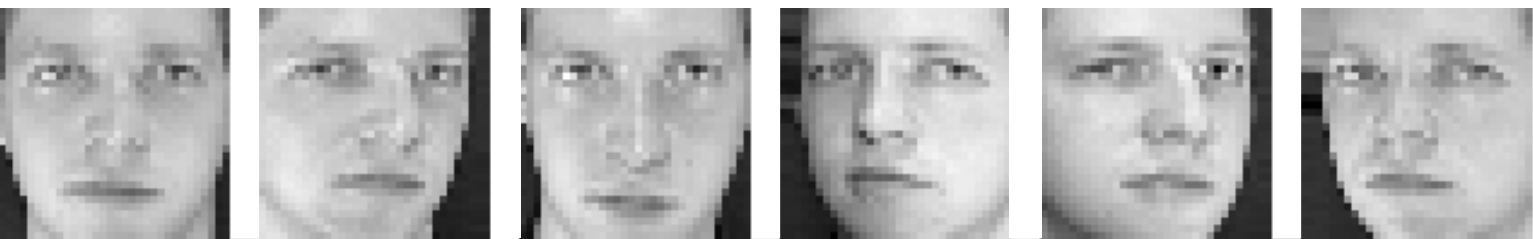}}
\subfigure[Class 3]{
\includegraphics[width=0.46\textwidth]{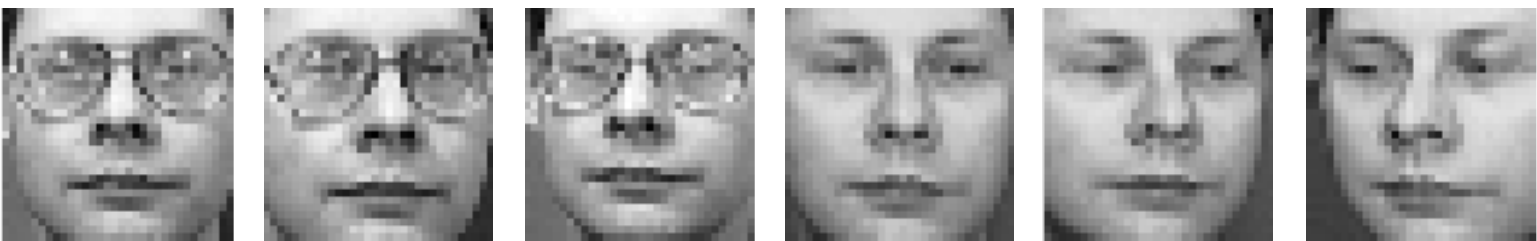}}
\end{center}
\caption{Samples from ORL dataset. Eye features are discriminative for class 1, and with less importances for class 3.}
\label{orlsample}
\end{figure}

NMF represents the non-negative data matrix with the linear combination of basis vectors. The coefficient matrix contains the low-dimensional representation of the original data. Since the coefficient matrix and basis vectors are constrained to be non-negative, NMF allows only additive operations. In this way, the part-of-whole interpretation is guaranteed, which complies with the human perception. Despite its advantages, NMF has several intrinsic limitations. Firstly, NMF neglects the local data structure. Supposing the samples are from $c$ classes, NMF represents the samples with only $c$ basis vectors. This strategy is unsuitable for the samples with non-Gaussian distributions. Secondly, NMF cannot distinguish the importances of the original features. In real-world tasks, some features should be emphasized~\cite{liufeature}. As shown in Fig.~\ref{orlsample} (a), the features of eyes are more discriminative than skins. Moreover, due to the diversity of samples, the importance of the same feature may vary across classes. In Fig.~\ref{orlsample} (b), the importance of the eyes decreases because the man wears a glass in some images. Since NMF treats all the features equally, the learned representation lacks discriminability and contains irrelevant features.

In recent years, many NMF based techniques have been proposed. To capture the local data structure, Cai et al.~\cite{gnmf} and Huang et al.~\cite{rmnmf} introduced the graph regularization term into the objective of NMF, and made the similar data points to be with consistent coefficient vectors. Gao et al.~\cite{lcnmf} and Chen et al.~\cite{chenconcept} exploited the local data relationship by introducing more local basis vectors. Han et al.~\cite{ongr} factorized the data graph to preserve the local manifold. Huang et al.~\cite{nmfan} proposed to learn the data relationship adaptively during matrix factorization. Liu et al.~\cite{lcf} employed local coordinate learning to constrain the basis vectors to be close to the samples. The above NMF variants solve the first problem well, and they are able to handle the data with various structures. However, the importances of features remains to be neglected in the literature. Some researchers~\cite{embed24,embed,lefeigraph,lefeigraph2,chenneru,ldanmf} projected the data into the subspace, and performed matrix factorization on the transformed data. This strategy alleviates the effect of irrelevant features, but fails to retain the original feature attributes. Importantly, all the samples are projected by one single transformation matrix, which may be insufficient to learn satisfactory representations.

To mitigate the above problems, we present the Feature weighted Non-negative Matrix Factorization (FNMF) method in this paper. The proposed model performs feature weighting and matrix factorization iteratively without any prior knowledge: 1) the weights of features are learned automatically according to their importance on matrix factorization; 2) the basis vectors and coefficient matrix are updated for the samples with weighted features. In addition, the graph regularization term is employed to capture the data manifold. The main contribution of this paper is summarized as follows.

\begin{itemize}
\item[1)] The feature weighting mechanism is introduced to distinguish the features with different importances. In this way, the irrelevant features are reduced and the informative ones are emphasized.

\item[2)] Considering the diversity of samples, each sample is associated with multiple feature weighting components to get the comprehensive representation. Data graph is also used to make the representation coherent in local neighborhoods.

\item[3)] An effective algorithm is designed to optimize the proposed model with low computational cost. Its convergence is proved both theoretically and experimentally.
\end{itemize}

The remaining parts are organized as follows. In Section~\ref{section_related}, the related works are revisited. In Section~\ref{section_fnmf}, the proposed model and the corresponding optimization algorithm are introduced.  In Section~\ref{section_exper}, experiments results are given to demonstrate the effectiveness. Section~\ref{section_conclusions} concludes this paper. 

\textbf{Notations:} throughout this paper, the matrices and vectors are written in uppercase and lowercase respectively. Given the matrix $\mathbf{A}$, the $(i,k)$-th element is denoted as $a_{ik}$. The $i$-th row and column are denoted as $\mathbf{a}_{i,:}$ and $\mathbf{a}_i$ respectively. $\mathbf{A}^T$ represents the transpose of $\mathbf{A}$. $\rm Tr()$ represents the trace operator. $\mathbf{I}$ indicates the identity matrix.

\section{Related Work}
\label{section_related}
In this section, we first briefly review the classical NMF~\cite{nmf} and then discuss some representative variants of NMF.

\subsection{Non-negative Matrix Factorization Revisited}
Convert each sample into a $d$ dimensional column vector, and denote the data matrix as $\mathbf{X} = [\mathbf{x}_1,\mathbf{x}_2,\cdots,\mathbf{x}_n]\in \mathbb{R}^{d\times n}$, where $n$ denotes the number of samples. NMF approximates $\mathbf{X}$ with the product of two non-negative matrices $\mathbf{U}\in \mathbb{R}^{d\times c}$ and $\mathbf{V}\in \mathbb{R}^{n\times c}$. Each column in $\mathbf{U}$ is a basis vector, and the $i$-th row of $\mathbf{V}$ is the low-dimensional representation of sample $\mathbf{x}_i$. Taking the least square error as the loss function, the objective of NMF is formulated as
\begin{equation}
\mathop {\min }\limits_{\mathbf{U} \ge 0,\mathbf{V} \ge 0} \sum\limits_{i = 1}^n {||{\mathbf{x}_i} - \mathbf{Uv}_{i,:}^T||_2^2},
\label{nmfobj}
\end{equation}
where $||\cdot||_2$ is the $\ell _2$ norm. The model is solved with the following multiplicative updating rules:
\begin{equation}
\begin{split}
{u_{ik}} \leftarrow {u_{ik}}\frac{{{{(\mathbf{XV})}_{ik}}}}{{{{(\mathbf{U}{\mathbf{V}^T}\mathbf{V})}_{ik}}}},\\
{v_{jk}} \leftarrow {v_{jk}}\frac{{{{({\mathbf{X}^T}\mathbf{U})}_{jk}}}}{{{{(\mathbf{V}{\mathbf{U}^T}\mathbf{U})}_{jk}}}}.
\end{split}
\end{equation}

From problem~(\ref{nmfobj}), we can see that NMF just focuses on the global reconstruction of the original data, and fails to preserve the local relationship. Meanwhile, all the features are concentrated together directly, so the irrelevant features affect the learned representation inevitably.

\subsection{Variants of NMF}
To learn the more effective data representation, many researches have been conducted to improve NMF.

Motivated by spectral clustering, Cai et al.~\cite{gnmf} encoded the manifold structure with a data graph to incorporate the geometry information. Kong et al.~\cite{l21nmf} replaced the Frobenious norm with the $\ell _{2,1}$ norm to improve the robustness. Huang et al.~\cite{rmnmf} and Chen et al.~\cite{chenrse} integrated the $\ell _{2,1}$ norm NMF with the local structure exploration. Ding et al.~\cite{dingconvex} used the convex combinations of the data points as basis vectors. They also proposed the semi-NMF to process the data negative values. Du et al.~\cite{correntropy} proposed the correntropy induced NMF to deal with the non-Gaussian outliers. Gao et al.~\cite{lcnmf} learned the structured bipartite graph with multiple local centroids to capture the neighboring relationship. Li et al.~\cite{dongcui} employed the low rank representation to perceive the underlying discriminant features. Inspired by the progress on graph clustering~\cite{can,clr,chenaaai}, Huang et al.~\cite{nmfan} found the neighbors of each sample adaptively, and used them to learn the optimal data graph. Wang et al.~\cite{jing} extended NMF to muti-view data representation. Wang et al.~\cite{hxnmf} learned the bi-stochastic data graph with a robust formulation. There are also many other techniques~\cite{l1nmf,sym,dingequivalence,missing1,xiaofeigraph}, which improve NMF from different perspectives skillfully. However, the above algorithms follow the assumption that the importances of different features are equal.

In order to reduce the irrelevant features, some methods proposed to perform feature learning and matrix simultaneously. Zhang et al.~\cite{lefeigraph} incorporated PCA into the NMF framework to learn the features within the subspace, and used them to guide the factorization procedure. Zhao et al.~\cite{lefeigraph2} performed dual matrix factorization within the original and projected data spaces jointly. Meanwhile, orthogonal constraint is imposed on the coefficient matrix to obtain a better interpretation. Zhang et al.~\cite{online} made use of the data labels to calculate the within- and between-class scatters, and utilized them to update the basis vectors. Yuan et al.~\cite{embed24} introduced the non-negative symmetric transformation matrix to find the optimal subspace. Belachew and Buono~\cite{embed} proposed the embedded projective NMF, which combines alternating least squares algorithm and multiplicative updating rules to accelerate the convergence. Chen et al.~\cite{chenneru} projected the samples with a structural sparse transformation matrix. Li et al.~\cite{ldanmf} bridged the connection between NMF and linear discriminant analysis, and learned the desired subspace for matrix approximation. All of these methods create new features to find the basis vectors, but the original data characteristics may be corrupted. Besides, it is exhausting to decide an appropriate dimension of the subspace.

\section{Feature Weighted Non-negative Matrix Factorization}
\label{section_fnmf}
In this section, we describe the Feature Weighted Non-negative Matrix Factorization (FNMF) method. First, the formulation of FNMF is proposed. Then, the optimization algorithm is designed to solve the proposed problem and the convergence analysis is provided.

\begin{figure}
\begin{center}
\includegraphics[width=0.48\textwidth]{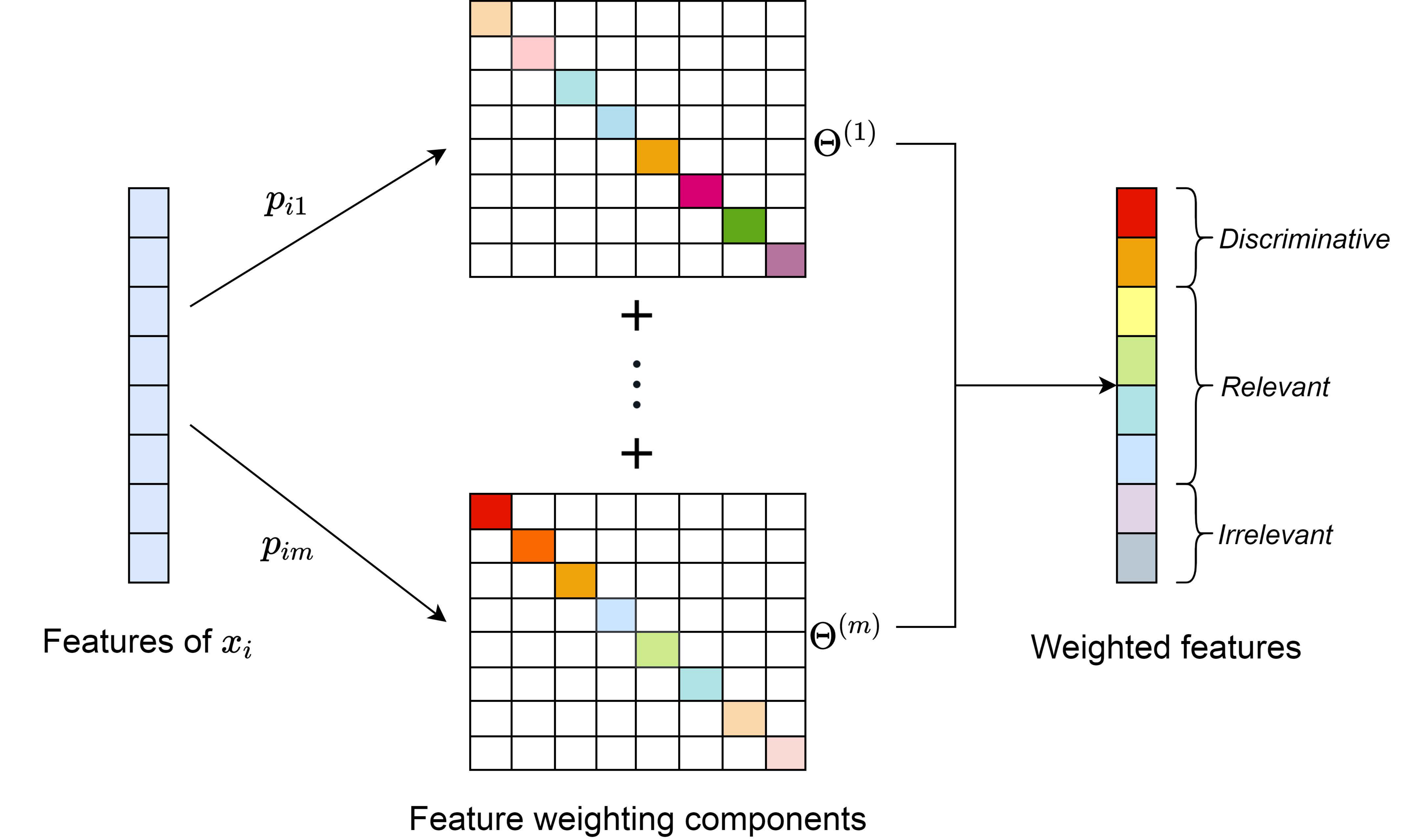}
\end{center}
\caption{Feature weighting with multiple components.}
\label{flow}
\end{figure}
\subsection{Methodology}
The exploration of informative features is crucial for learning an effective data representation. We propose to learn the weights of features automatically, without any prior knowledge about the subspace dimension and data label. Firstly, by introducing a feature weighting matrix, the objective is written as
\begin{equation}
\begin{split}
&\mathop {\min }\limits_{\mathbf{ \Theta,U,V}} \sum\limits_{i = 1}^n {||\mathbf{\Theta} {\mathbf{x}_i} - \mathbf{Uv}_{i,:}^T||_2^2}+\beta {\rm Tr}(\mathbf{V}^T\mathbf{LV}), \\
&{\rm s.t.} \ \mathbf{\Theta}  = {\rm diag}(\theta ), {\theta _k} \ge 0,\sum\limits_{k = 1}^d {{\theta _k}}  = 1,\\
& \quad \ \ \mathbf{U} \ge 0,\mathbf{V} \ge 0,
\end{split}
\end{equation}
where the diagonal matrix $\mathbf{\Theta} \in \mathbb{R}^{d\times d}$ assigns each feature with a weight, and $\theta _k$ measures the importance of the $k$-th feature. $\beta$ is the regularization parameter, and $\mathbf{L} \in \mathbb{R}^{n\times n}$ is the Laplacian matrix of the predefined similarity graph. The second term makes the samples with small distances to share similar representations. 

Due to the complexity of the real-world data, the significance of a specific feature may vary for different samples, which can be seen from the comparison of Fig.~\ref{orlsample} (a) and (b). Therefore, we propose to incorporate multiple feature weighting components to capture the diversity, and the FNMF model is formulated as
\begin{equation}
\begin{split}
&\mathop {\min }\limits_{\mathbf{\Theta, P,U,V}} \ \sum\limits_{i = 1}^n {\sum\limits_{j = 1}^m {||{\mathbf{\Theta} ^{(j)}}{\mathbf{x}_i} - \mathbf{Uv}_{i,:}^T||_2^2p_{ij}^2} }  + \lambda \sum\limits_{j \ne l} {{\rm Tr}({\mathbf{\Theta} ^{(j)}}{\mathbf{\Theta} ^{(l)}})}\\
 & \qquad \quad \ \ + \beta {\rm Tr}({\mathbf{V}^T}\mathbf{LV}),\\
&{\rm s.t.} \  \mathbf{\Theta} ^{(j)}  = {\rm diag}(\theta ^{(j)}),
{\theta ^{(j)} _k}\ge 0,  \sum\limits_{k = 1}^d {{\theta ^{(j)}_k}}  = 1, \mathbf{U} \ge 0, \mathbf{V} \ge 0,\\
& \quad \ \  p_{ij} \ge 0, \sum\limits_{j=1}^m{p_{ij}}=1, 
\end{split}
\label{finalobj}
\end{equation}
where $\mathbf{\Theta} ^{(j)}$ is the $j$-th feature weighting component, $p_{ij}$ indicates the probability that the $\mathbf{x}_i$ is associated with $\mathbf{\Theta} ^{(j)}$, $m$ counts the number of components, $\lambda$ is the regularization parameter. In problem~(\ref{finalobj}), $p_{ij}$ and $\mathbf{\Theta} ^{j}$ are adjusted adaptively according to the current $\mathbf{U}$ and $\mathbf{V}$. The samples are assigned with different feature weighting components, and the diversity of the components is enforced by the second term. For the sample $\mathbf{x}_i$, its $k$-th feature will be paid more attention if both $p_{ij}$ and $\theta _k^{(j)}$ are large.

\subsection{Optimization Algorithm}
Problem~(\ref{finalobj}) contains four variables, so we decompose it into four sub-problems, and solve them iteratively.

\textbf{When updating} $\mathbf{\Theta}$, problem~(\ref{finalobj}) becomes
\begin{equation}
\begin{split}
&\mathop {\min }\limits_{\mathbf{\Theta}} \ \sum\limits_{i = 1}^n {\sum\limits_{j = 1}^m {||{\mathbf{\Theta} ^{(j)}}{\mathbf{x}_i} - \mathbf{Uv}_{i,:}^T||_2^2p_{ij}^2} }  + \lambda \sum\limits_{j \ne l} {{\rm Tr}({\mathbf{\Theta} ^{(j)}}{\mathbf{\Theta} ^{(l)}})},\\
&{\rm s.t.} \  \mathbf{\Theta} ^{(j)}  = {\rm diag}(\theta ^{(j)}),
{\theta ^{(j)} _k}\ge 0,  \sum\limits_{k = 1}^d {{\theta ^{(j)}_k}}  = 1.
\end{split}
\end{equation}
The problem can be optimized for each $\mathbf{\Theta} ^{(j)}$ as
\begin{equation}
\begin{split}
&\mathop {\min }\limits_{\mathbf{\Theta} ^{(j)}} \ \sum\limits_{i = 1}^n {||{\mathbf{\Theta} ^{(j)}}{\mathbf{x}_i} - \mathbf{Uv}_{i,:}^T||_2^2p_{ij}^2}  + \lambda {\rm Tr}({\mathbf{\Theta} ^{(j)}}\sum\limits_{l \ne j} {{\mathbf{\Theta} ^{(l)}}} ),\\
&{\rm s.t.} \  \mathbf{\Theta} ^{(j)}  = {\rm diag}(\theta ^{(j)}),
{\theta ^{(j)} _k}\ge 0,  \sum\limits_{k = 1}^d {{\theta ^{(j)}_k}}  = 1, 
\end{split}
\end{equation}
which is equivalent to
\begin{equation}
\begin{split}
&\mathop {\min }\limits_{\mathbf{\Theta} ^{(j)}} {\rm Tr}({\mathbf{\Theta} ^{(j)}}\sum\limits_{i = 1}^n {p_{ij}^2{\mathbf{x}_i}\mathbf{x}_i^T} {\mathbf{\Theta} ^{(j)}}) + \lambda {\rm Tr}({\mathbf{\Theta} ^{(j)}}\sum\limits_{l \ne j} {{\mathbf{\Theta} ^{(l)}}} ) \\
&\qquad - 2{\rm Tr}({\mathbf{\Theta} ^{(j)}}\sum\limits_{i = 1}^n {p_{ij}^2{\mathbf{x}_i}{\mathbf{v}_{i,:}\mathbf{U}^T}} ),\\
&\ \ {\rm s.t.} \  \mathbf{\Theta} ^{(j)}  = {\rm diag}(\theta ^{(j)}),
{\theta ^{(j)} _k}\ge 0,  \sum\limits_{k = 1}^d {{\theta ^{(j)}_k}}  = 1. 
\end{split}
\end{equation}
Note that $\mathbf{\Theta} ^{(j)}$ is a diagonal matrix. Denoting the $k$-th diagonal elements of $\sum\limits_{i = 1}^n {p_{ij}^2{\mathbf{x}_i}\mathbf{x}_i^T} {\mathbf{\Theta} ^{(j)}}$ and $\lambda \sum\limits_{l \ne j} {{\mathbf{\Theta} ^{(l)}}}  - 2\sum\limits_{i = 1}^n {p_{ij}^2{\mathbf{x}_i}{{\mathbf{v}_{i,:}}\mathbf{U}^T}}$ as $a_k$ and $b_k$ respectively, the problem is converted into
\begin{equation}
\begin{split}
&\mathop {\min }\limits_{{\theta ^{(j)}}} \  \sum\limits_{k = 1}^d {\theta _k^{{{(j)}^2}}{a_k} + \theta _k^{(j)}{b_k}} ,\\
&{\rm{s}}.{\rm{t}}.\ \ {\theta ^{(j)} _k}\ge 0,\sum\limits_{k = 1}^d {\theta _k^{(j)}}  = 1.
\end{split}
\label{updatetheta}
\end{equation}
which has the close form solution and can be optimized by an efficient approach~\cite{clr}.

\textbf{When Updating} $\mathbf{P}$, the sub-problem is
\begin{equation}
\begin{split}
&\mathop {\min }\limits_{\mathbf{P}} \ \sum\limits_{i = 1}^n {\sum\limits_{j = 1}^m {||{\mathbf{\Theta} ^{(j)}}{\mathbf{x}_i} - \mathbf{Uv}_{i,:}^T||_2^2p_{ij}^2} } ,\\
&{\rm s.t.} \  p_{ij} \ge 0, \sum\limits_{j=1}^m{p_{ij}}=1. 
\end{split}
\end{equation}
Denoting a diagonal matrix $\mathbf{Q}\in\mathbb{R}^{m\times m}$ with $ q_{j}=\sum\limits_{i = 1}^n {||{\mathbf{\Theta} ^{(j)}}{\mathbf{x}_i} - \mathbf{Uv}_{i,:}^T||_2^2}$, we have:
\begin{equation}
\begin{split}
&\mathop {\min }\limits_{\mathbf{p}_{i,:}} \ {\mathbf{p}_{i,:}^T\mathbf{Qp}_{i,:}}, \\
&{\rm s.t.} \  \mathbf{p}_{i,:} \ge 0, \mathbf{p}_{i,:}^T\mathbf{1}=1, 
\end{split}
\end{equation}
where $\mathbf{1}$ is a column vector with all its elements as 1.
Removing the constraint $p_{ij} \ge 0$, the Lagrangian function is
\begin{equation}
{\cal L}(p_{i,:},\eta) =  \mathbf{p}_{i,:}^T\mathbf{Qp}_{i,:}+\eta (1-\mathbf{p}_{i,:}^T\mathbf{1}),
\end{equation}
where the scalar $\eta$ is the Lagrangian multiplier. Let $\frac{{\partial {\cal L}({\mathbf{p}_{i,:}},\eta )}}{{\partial {\mathbf{p}_{i,:}}}}$ to be zero, we arrive at
\begin{equation}
2\mathbf{Qp}_{i,:}-\eta \mathbf{1} = 0,
\end{equation}
which further yields to
\begin{equation}
p_{ij} = \frac{\eta}{2q_{j}}.
\end{equation}
According to the constraint $\mathbf{p}_{i,:}^T\mathbf{1}=1$, it is easy to calculate the value of $\eta$. Together with the definition of $\mathbf{Q}$, we have
\begin{equation}
p_{ij} = \frac{1}{\sum\limits_{i = 1}^n {||{\mathbf{\Theta} ^{(j)}}{\mathbf{x}_i} - \mathbf{Uv}_{i,:}^T||_2^2}}/ (\sum \limits _{l=1}^m \frac{1}{ \sum\limits_{i = 1}^n {||{\mathbf{\Theta} ^{(l)}}{\mathbf{x}_i} - \mathbf{Uv}_{i,:}^T||_2^2}}),
\label{updatep}
\end{equation}
which satisfies the constraint $p_{ij} \ge 0$ definitely.

\textbf{When updating} $\mathbf{U}$, the objective is transformed into
\begin{equation}
\begin{split}
\mathop {\min }\limits_{\mathbf{U}\ge 0} \ \sum\limits_{i = 1}^n {\sum\limits_{j = 1}^m {||{\mathbf{\Theta} ^{(j)}}{\mathbf{x}_i} - \mathbf{Uv}_{i,:}^T||_2^2p_{ij}^2} },
\end{split}
\end{equation}
Removing the irrelevant term, the problem is reformulated as
\begin{equation}
\begin{split}
\mathop {\min }\limits_{\mathbf{U}\ge 0} \ &{\rm Tr}({\mathbf{U}^T}\mathbf{U}\sum\limits_{i = 1}^n {\sum\limits_{j = 1}^m {p_{ij}^2\mathbf{v}_{i,:}^T{\mathbf{v}_{i,:}}} } ) \\
&- 2{\rm Tr}(\mathbf{U}\sum\limits_{i = 1}^n {\sum\limits_{j = 1}^m {p_{ij}^2\mathbf{v}_{i,:}^T\mathbf{x}_i^T{\mathbf{\Theta} ^{{{(j)}^T}}}} } ).
\end{split}
\end{equation}
Accordingly, the Lagrangian function is
\begin{equation}
\begin{split}
{\cal L}(\mathbf{U},\mathbf{\Phi} ) =& {\rm Tr}({\mathbf{U}^T}\mathbf{U}\sum\limits_{i = 1}^n {\sum\limits_{j = 1}^m {\mathbf{p}_{ij}^2\mathbf{v}_{i,:}^T{\mathbf{v}_{i,:}}} } ) \\
&- 2{\rm Tr}(\mathbf{U}\sum\limits_{i = 1}^n {\sum\limits_{j = 1}^m {p_{ij}^2\mathbf{v}_{i,:}^Tx_i^T{\mathbf{\Theta} ^{{{(j)}^T}}}} } ) + {\rm Tr}(\mathbf{\Phi} {\mathbf{U}^T}),
\end{split}
\end{equation}
where $\mathbf{\Phi} \in \mathbb{R}^{c\times c}$ is the Lagrangian multiplier. ${\cal L}(\mathbf{U,\Phi} )$ is convex w.r.t. $\mathbf{U}$, so the optimal $\mathbf{U}$ satisfies
\begin{equation}
\begin{split}
&\frac{{\partial {\cal L}(\mathbf{U,\Phi} )}}{{\partial U}} \\
&= 2\mathbf{U}\sum\limits_{i = 1}^n {\sum\limits_{j = 1}^m {p_{ij}^2\mathbf{v}_{i,:}^T{\mathbf{v}_{i,:}}} }  - 2\sum\limits_{i = 1}^n {\sum\limits_{j = 1}^m {p_{ij}^2{\mathbf{\Theta} ^{(j)}}{\mathbf{x}_i}{\mathbf{v}_{i,:}}} }  + \mathbf{\Phi}  \\
&= 0.
\end{split}
\end{equation}
According to the KKT conditions,we have $\phi _{ik} u_{ik}=0$, which leads to
\begin{equation}
\begin{split}
{(\mathbf{U}\sum\limits_{i = 1}^n {\sum\limits_{j = 1}^m {p_{ij}^2\mathbf{v}_{i,:}^T{\mathbf{v}_{i,:}}} }  - \sum\limits_{i = 1}^n {\sum\limits_{j = 1}^m {p_{ij}^2{\mathbf{\Theta} ^{(j)}}{\mathbf{x}_i}{\mathbf{v}_{i,:}}} } )_{ik}}u_{ik}^2 &= \phi _{ik} u^2_{ik} \\
&= 0.
\end{split}
\end{equation}
Therefore, the updating rule of $\mathbf{U}$ is
\begin{equation}
{u_{ik}} \leftarrow {u_{ik}}\sqrt {\frac{{{{(\sum\limits_{i = 1}^n {\sum\limits_{j = 1}^m {p_{ij}^2{\mathbf{\Theta} ^{(j)}}{\mathbf{x}_i}{\mathbf{v}_{i,:}}} } )}_{ik}}}}{{{{(\mathbf{U}\sum\limits_{i = 1}^n {\sum\limits_{j = 1}^m {p_{ij}^2\mathbf{v}_{i,:}^T{\mathbf{v}_{i,:}}} } )}_{ik}}}}} .
\label{updateu}
\end{equation}
At convergence, $\mathbf{U}$ satisfies the condition $\frac{\partial {\cal L}(\mathbf{U,\Phi} )}{\partial \mathbf{U}}=0$.

\textbf{When updating} $\mathbf{V}$, the objective becomes
\begin{equation}
\mathop {\min }\limits_{\mathbf{V}\ge 0} \ \sum\limits_{i = 1}^n {\sum\limits_{j = 1}^m {||{\mathbf{\Theta} ^{(j)}}{\mathbf{x}_i} - \mathbf{Uv}_{i,:}^T||_2^2p_{ij}^2} }   + \beta {\rm Tr}({\mathbf{V}^T}\mathbf{LV}).
\end{equation}
Denoting the similarity graph as $\mathbf{S}\in\mathbb{R}^{n\times n}$, where the diagonal elements are set as zero, we know
\begin{equation}
\min \limits_\mathbf{V} {\rm Tr}({\mathbf{V}^T}\mathbf{LV}) = \min \limits_\mathbf{V}\sum \limits _{i\ne r} {||\mathbf{v}_{i,:}-\mathbf{v}_{r,:}||_2^2s_{ir}},
\end{equation}
so we can solve each $\mathbf{v}_{i,:}$ independently
\begin{equation}
\begin{split}
\mathop {\min }\limits_{\mathbf{v}_{i,:}\ge 0} \ &\sum\limits_{j = 1}^m {(p_{ij}^2{\mathbf{v}_{i,:}}{\mathbf{U}^T}\mathbf{Uv}_{i,:}^T - 2p_{ij}^2{\mathbf{v}_{i,:}}{\mathbf{U}^T}{\mathbf{\Theta} ^{(j)}}{\mathbf{x}_i})} \\
 &+ \beta \sum\limits_{r\ne i} {{s_{ir}}} {\mathbf{v}_{i,:}}\mathbf{v}_{i,:}^T - 2\beta \sum\limits_{r\ne i} {{s_{ir}}} {\mathbf{v}_{i,:}}\mathbf{v}_{r,:}^T.
\end{split}
\end{equation}
The Lagrangian function is
\begin{equation}
\begin{split}
{\cal L}({\mathbf{v}_{i,:}},\varphi ) =&\sum\limits_{j = 1}^m {(p_{ij}^2{\mathbf{v}_{i,:}}{\mathbf{U}^T}\mathbf{Uv}_{i,:}^T - 2p_{ij}^2{\mathbf{v}_{i,:}}{\mathbf{U}^T}{\mathbf{\Theta} ^{(j)}}{\mathbf{x}_i})} \\
 &+ \beta \sum\limits_{r\ne i} {{s_{ir}}} {\mathbf{v}_{i,:}}\mathbf{v}_{i,:}^T - 2\beta \sum\limits_{r\ne i} {{s_{ir}}} {\mathbf{v}_{i,:}}\mathbf{v}_{r,:}^T + \varphi \mathbf{v}_{i,:}^T.
\label{lv}
\end{split}
\end{equation}
$\varphi \in \mathbb{R}^{1\times n}$
is the Lagrangian multiplier. Let $\frac{\partial {\cal L}(\mathbf{v}_{i,:},\varphi)}{\partial \mathbf{v}_{i,:}}$ to be zero, we have
\begin{equation}
\begin{split}
\frac{\partial {\cal L}(\mathbf{v}_{i,:},\varphi)}{\partial \mathbf{v}_{i,:}}=&2{\mathbf{v}_{i,:}}({\mathbf{U}^T}\mathbf{U}\sum\limits_{j = 1}^m {p_{ij}^2}  + \beta \sum\limits_{r\ne i} {{s_{ir}}} \mathbf{I}) \\
&- 2(\sum\limits_{j = 1}^m {p_{ij}^2\mathbf{x}_i^T{\mathbf{\Theta} ^{{{(j)}^T}}}U}  + \beta \sum\limits_{r\ne i} {{s_{ir}}} {\mathbf{v}_{r,:}}) + \varphi .
\end{split}
\end{equation}
With the KKT condition $\varphi _{k} \mathbf{v}_{ik}=0$, the updating rule of $\mathbf{V}$ is
\begin{equation}
{v_{ik}} \leftarrow {v_{ik}}\sqrt {\frac{{{{(\sum\limits_{j = 1}^m {p_{ij}^2\mathbf{x}_i^T{\mathbf{\Theta} ^{{{(j)}^T}}}\mathbf{U}}  + \beta \sum\limits_{r\ne i} {{s_{ir}}} {\mathbf{v}_{r,:}})_k}}}}{{{{({\mathbf{v}_{i,:}}{\mathbf{U}^T}\mathbf{U}\sum\limits_{j = 1}^m {p_{ij}^2}  + \beta \sum\limits_{r\ne i} {{s_{ir}}{\mathbf{v}_{i,:}}} )_k}}}}} .
\label{updatev}
\end{equation}

The detailed algorithm to solve problem~(\ref{finalobj}) is outlined in Algorithm~\ref{alg1}. The computational cost to construct the data graph $\mathbf{S}$ is ${\cal O}(nnd)$. In each iteration, the complexity of updating $\mathbf{U}$, $\mathbf{\Theta}$ and $\mathbf{P}$ is ${\cal O}(mndc)$. Let $\mathbf{S}$ be a $K$ neighbors sparse graph, the time cost of updating $\mathbf{V}$ is ${\cal O}(mndc+mnKc)$. After $t$ iterations, the overall computational cost of Algorithm~\ref{alg1} is ${\cal}(nnd+mndc+mnKc)$. Compared with the graph-regularized NMF~\cite{gnmf}, FNMF has slightly higher cost due to the incorporation of multiple feature weighting components, but it converges very fast, which will be demonstrated in Section~\ref{section_exper}.

\begin{algorithm}
        \caption{Optimization algorithm of FMMF}
        \begin{algorithmic}[1]
        \Require Data matrix $\mathbf{X}$, class number $c$, parameter $\lambda$, $\beta$.
           \State Initializing $\Theta$, $\mathbf{P}$, $\mathbf{U}$ and $\mathbf{V}$.
           \State Constructing similarity graph $\mathbf{S}$.
           
        	\Repeat
			\State Updating $\mathbf{\Theta}$ by solving problem~(\ref{updatetheta}).
			\State Updating $\mathbf{P}$ with Eq.~(\ref{updatep}). 
			\State Updating $\mathbf{U}$ with Eq.~(\ref{updateu}).
			\State Updating $\mathbf{V}$ with Eq.~(\ref{updatev}).
			\Until{Convergence}
		\Ensure Optimal $\mathbf{U}$, $\mathbf{V}$.
        \end{algorithmic}
        \label{alg1}
    \end{algorithm}

\begin{figure*}
\begin{center}
\subfigure[First two dimensions]{
\includegraphics[width=0.18\textwidth]{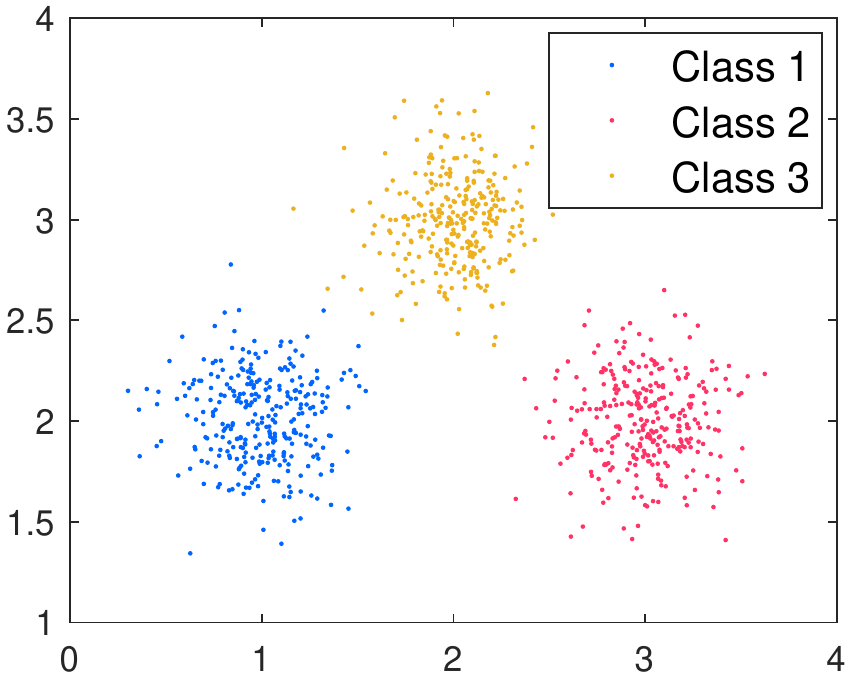}}
\subfigure[NMF]{
\includegraphics[width=0.18\textwidth]{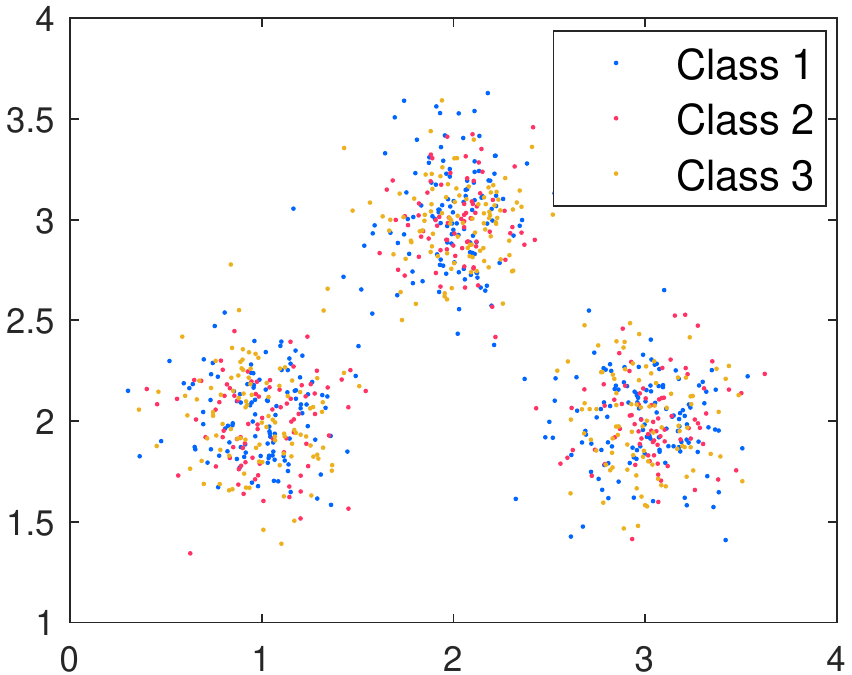}}
\subfigure[RNMF]{
\includegraphics[width=0.18\textwidth]{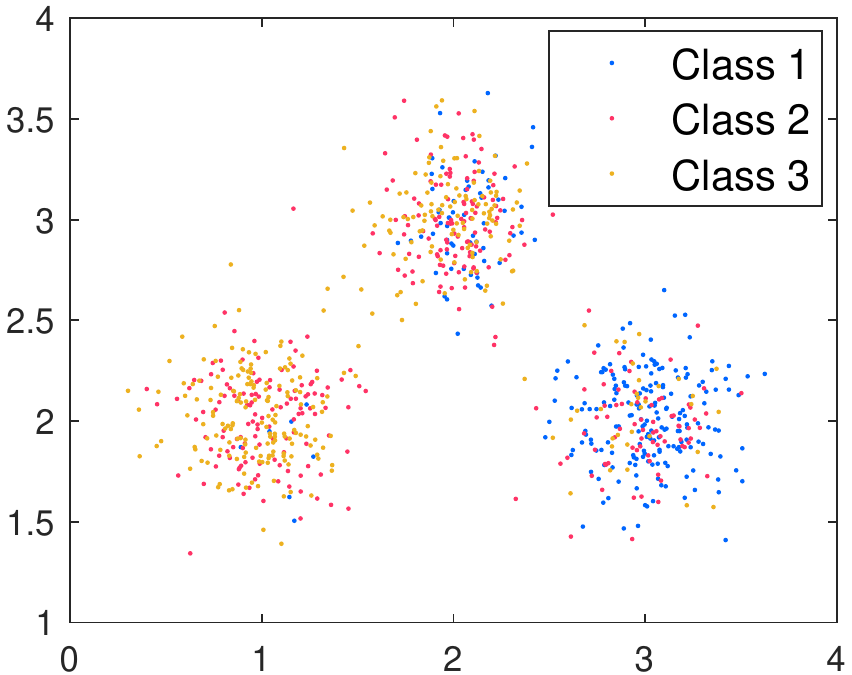}}
\subfigure[GNMF]{
\includegraphics[width=0.18\textwidth]{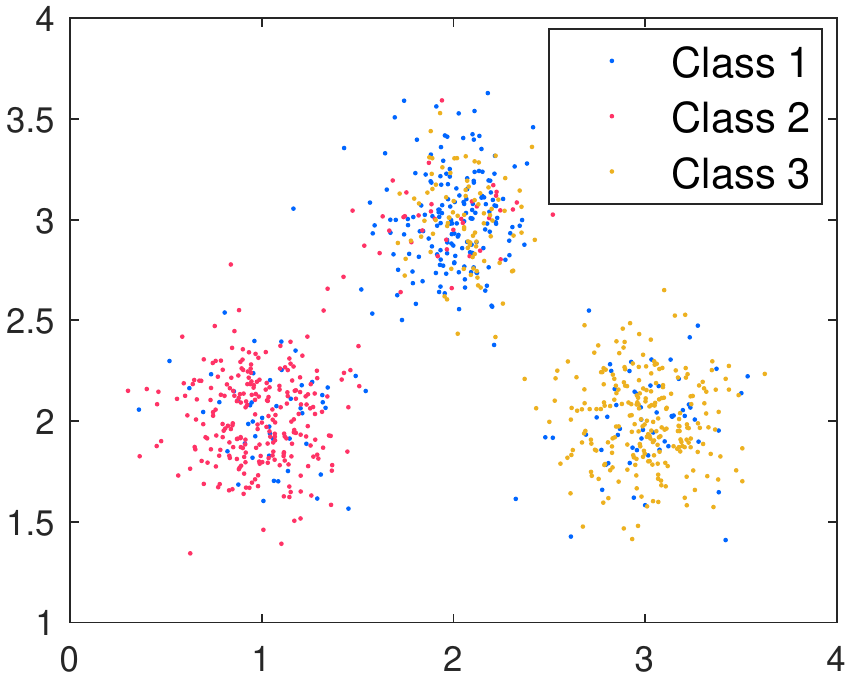}}
\subfigure[FNMF]{
\includegraphics[width=0.18\textwidth]{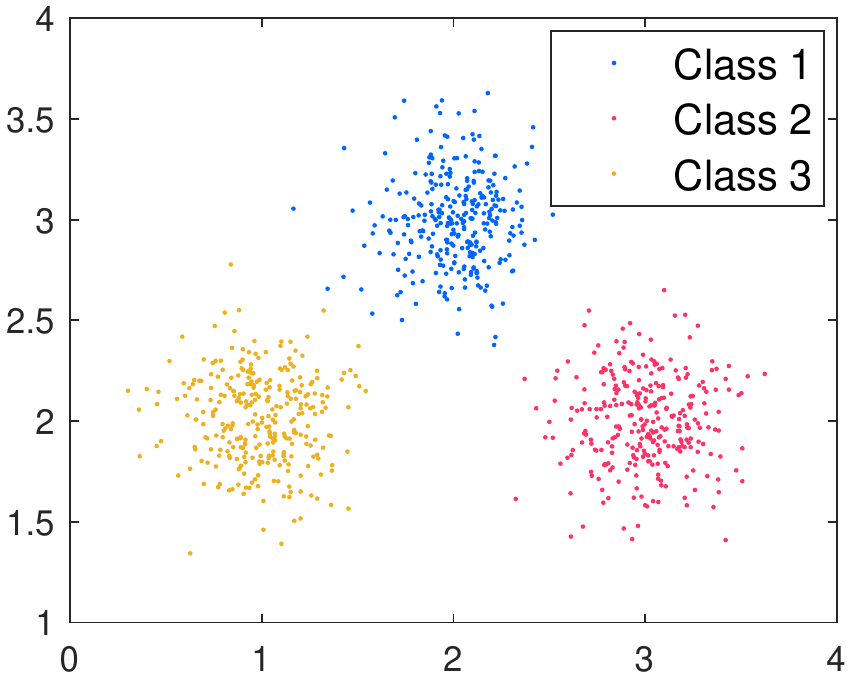}}
\end{center}
\caption{Results on the three-Gaussian toy dataset with five noisy dimensions.}
\label{toy1}
\end{figure*}

\subsection{Convergence Analysis}
During each iteration, the optimization of $\mathbf{\Theta}$ and $\mathbf{P}$ decreases the objective value monotonically because the solutions satisfy the KKT condition. $\mathbf{U}$ and $\mathbf{V}$ are updated in similar forms, so we mainly demonstrate the convergence of the updating rule~(\ref{updatev}).

Since the last term in Eq.~(\ref{lv}) is zero, the Lagrangian function can be rewritten as
\begin{equation}
\begin{split}
{\cal L}(\mathbf{v}_{i,:}) = &{\mathbf{v}_{i,:}}({\mathbf{U}^T}\mathbf{U}\sum\limits_{j = 1}^m {p_{ij}^2}  + \beta \sum\limits_{j = 1}^n {{s_{ij}\mathbf{I}}} )\mathbf{v}_{i,:}^T \\
&- 2(\sum\limits_{j = 1}^m {p_{ij}^2\mathbf{x}_i^T{\mathbf{\Theta} ^{{{(j)}^T}}}\mathbf{U}}  + \beta \sum\limits_{r\ne i} {{s_{ir}}} {\mathbf{v}_{r,:}})\mathbf{v}_{i,:}^T.
\end{split}
\end{equation}
According to Lee and Seung~\cite{nmf}, we have the following definition and lemma.
\begin{Definition}
${\cal G}(\mathbf{v}_{i,:},\mathbf{v}_{i,:}^{(t)})$ is the auxiliary function for ${\cal L}(\mathbf{v}_{i,:})$ if the following condition holds
\begin{equation}
{\cal G}(\mathbf{v}_{i,:},\mathbf{v}_{i,:}^{(t)})\ge {\cal L}(\mathbf{v}_{i,:}), {\cal G}(\mathbf{v}_{i,:},\mathbf{v}_{i,:})={\cal L}(\mathbf{v}_{i,:}).
\end{equation}
\end{Definition}
\begin{Lemma}
Given the auxiliary function ${\cal G}(\mathbf{v}_{i,:},\mathbf{v}_{i,:}^{(t)})$, ${\cal L}(\mathbf{v}_{i,:}^{(t+1)})\le {\cal L}(\mathbf{v}_{i,:}^{(t)})$ holds if $\mathbf{v}_{i,:}^{(t+1)}$ is the solution to
\begin{equation}
\min \limits _{\mathbf{v}_{i,:}} {\cal G}(\mathbf{v}_{i,:},\mathbf{v}_{i,:}^{(t)}).
\end{equation}
\label{lemma1}
\end{Lemma}
Therefore, the key steps are: 1) find the auxiliary function ${\cal G}(\mathbf{v}_{i,:},\mathbf{v}_{i,:}^{(t)})$ of ${\cal L}(\mathbf{v}_{i,:})$; 2) get the global minimum value of ${\cal G}(\mathbf{v}_{i,:},\mathbf{v}_{i,:}^{(t)})$. To prove the convergence, the following theorem is introduced.

\begin{theorem} 
${\cal L}(\mathbf{v}_{i,:})$ is non-increasing under the updating rule (\ref{updatev}).
\end{theorem}
\begin{proof}
As Ding et al.~\cite{dingconvex} demonstrated, for any $\mathbf{H} \in \mathbb{R}^{c\times c}$, $\mathbf{e} \in \mathbb{R}^{1\times c}$ and $\mathbf{z} \in \mathbb{R}^{1\times c}$, if $\mathbf{H}$ is symmetric, we have the following inequality
\begin{equation}
{\rm Tr}(\mathbf{e}^T\mathbf{eH})\le \sum\limits _{k=1}^c {\frac{(\mathbf{zH})_k e_{k}^2}{z_k}}.
\end{equation}
Therefore, it can be deduced that
\begin{equation}
\begin{split}
&{\mathbf{v}_{i,:}}({\mathbf{U}^T}\mathbf{U}\sum\limits_{j = 1}^m {p_{ij}^2}  + \beta \sum\limits_{r = 1}^n {{s_{ir}}\mathbf{I}} )\mathbf{v}_{i,:}^T \\
&= {\rm Tr}[\mathbf{v}_{i,:}^T{\mathbf{v}_{i,:}}({\mathbf{U}^T}\mathbf{U}\sum\limits_{j = 1}^m {p_{ij}^2}  + \beta \sum\limits_{r\ne i} {{s_{ir}}\mathbf{I}} )] \\
&\le \sum\limits_{k = 1}^c {\frac{{{{(\mathbf{v}_{i,:}^{(t)}{\mathbf{U}^T}\mathbf{U}\sum\limits_{j = 1}^m {p_{ij}^2}  + \beta \sum\limits_{r\ne i} {{s_{ir}\mathbf{v}_{i,:}^{(t)}}} )_k}}v_{ik}^2}}{v_{ik}^{(t)}}} ,
\end{split}
\end{equation}
where $v_{ik}^{(t)}$ is a scalar.
In addition, we have $\frac{v_{ik}}{v^{(t)}_{ik}}\ge 1+\log \frac{v_{ik}}{v^{(t)}_{ik}}$, which leads to
\begin{equation}
\begin{split}
&(\sum\limits_{j = 1}^m {p_{ij}^2\mathbf{x}_i^T{\mathbf{\Theta} ^{{{(j)}^T}}}\mathbf{U}}  + \beta \sum\limits_{r = 1}^n {{s_{ir}}} {\mathbf{v}_{r,:}})\mathbf{v}_{i,:}^T \\
&= \sum\limits_{k = 1}^c {(\sum\limits_{j = 1}^m {p_{ij}^2\mathbf{x}_i^T{\mathbf{\Theta} ^{{{(j)}^T}}}\mathbf{U}}  + \beta \sum\limits_{r\ne i} {{s_{ij}}} {\mathbf{v}_{r,:}})_k{v_{ik}}} \\
& \ge \sum\limits_{k = 1}^c {{{(\sum\limits_{j = 1}^m {p_{ij}^2\mathbf{x}_i^T{\mathbf{\Theta} ^{{{(j)}^T}}}\mathbf{U}}  + \beta \sum\limits_{r\ne i} {{s_{ir}}} {\mathbf{v}_{r,:}})_k}}{v^{(t)}_{ik}}(1 + \log\frac{{{v_{ik}}}}{v^{(t)}_{ik}})} .
\end{split}
\end{equation}
Combining the above bounds, the auxiliary function of ${\cal L}(\mathbf{v}_{i,:})$ is 
\begin{equation}
\begin{split}
&{\cal G}(\mathbf{v}_{i,:},\mathbf{v}_{i,:}^{(t)})=\sum\limits_{k = 1}^c {\frac{{{{({\mathbf{v}^{(t)}_{i,:}}{U^T}U\sum\limits_{j = 1}^m {p_{ij}^2}  +  \beta\sum\limits_{r\ne i} {{s_{ir}}{\mathbf{v}^{(t)}_{i,:}}} )_k}}v_{ik}^2}}{{{v^{(t)}_{ik}}}}}\\
&-2\sum\limits_{k = 1}^c {{{(\sum\limits_{j = 1}^m {p_{ij}^2\mathbf{x}_i^T{\mathbf{\Theta} ^{{{(j)}^T}}}\mathbf{U}}  + \beta \sum\limits_{r\ne i} {{s_{ir}}} {\mathbf{v}_{r,:}})_k}}{v^{(t)}_{ik}}(1 + \log\frac{{{v_{ik}}}}{v^{(t)}_{ik}})}.
\end{split}
\end{equation}
The first-order derivative of ${\cal G}(\mathbf{v}_{i,:},\mathbf{v}_{i,:}^{(t)})$ w.r.t. $v_{ik}$ is 
\begin{equation}
\begin{split}
\frac{\partial {\cal G}(\mathbf{v}_{i,:},\mathbf{v}_{i,:}^{(t)})}{\partial v_{ik}} =& \frac{{2{({\mathbf{v}^{(t)}_{i,:}}{\mathbf{U}^T}\mathbf{U}\sum\limits_{j = 1}^m {p_{ij}^2}  + \beta \sum\limits_{r\ne i} {{s_{ij}}\mathbf{v}^{(t)}_{i,:} )_k}}{v_{ik}}}}{{{v^{(t)}_{ik}}}}\\
& - \frac{{2{{(\sum\limits_{j = 1}^m {p_{ij}^2\mathbf{x}_i^T{\mathbf{\Theta} ^{{{(j)}^T}}}\mathbf{U}}  + \beta \sum\limits_{r\ne i} {{s_{ir}}} {\mathbf{v}_{r,:}})_k}}{v^{(t)}_{ik}}}}{{{v_{ik}}}},
\end{split}
\end{equation}
and the $k$-th diagonal element of the Hessian matrix is
\begin{equation}
\begin{split}
\frac{\partial ^2 {\cal G}(\mathbf{v}_{i,:},\mathbf{v}_{i,:}^{(t)})}{\partial v_{ik} \partial v_{ik}}=& \frac{{2{({\mathbf{v}^{(t)}_{i,:}}{\mathbf{U}^T}\mathbf{U}\sum\limits_{j = 1}^m {p_{ij}^2}  + \beta \sum\limits_{r\ne i} {{s_{ij}}\mathbf{v}^{(t)}_{i,:} )_k}}}}{{{v^{(t)}_{ik}}}}\\
&  + \frac{{2{{(\sum\limits_{j = 1}^m {p_{ij}^2\mathbf{x}_i^T{\mathbf{\Theta} ^{{{(j)}^T}}}\mathbf{U}}  + \beta \sum\limits_{r\ne i} {{s_{ir}}} {\mathbf{v}_{r,:}})_k}}{v^{(t)}_{ik}}}}{{{v^2_{ik}}}}.
\end{split}
\end{equation}
Since the Hessian matrix is semi-positive definite, ${\cal G}(\mathbf{v}_{i,:},\mathbf{v}_{i,:}^{(t)})$ is convex on $\mathbf{v}_{i,:}$. Therefore, the global optimal solution $\mathbf{v}_{i,:}^{(t+1)}$ to $\min \limits _{\mathbf{v}_{i,:}} {\cal G}(\mathbf{v}_{i,:},\mathbf{v}_{i,:}^{(t)})$ can be computed by setting the first-order derivative to zero:
\begin{equation}
{v^{(t+1)}_{ik}}= {v^{(t)}_{ik}}\sqrt {\frac{{{{(\sum\limits_{j = 1}^m {p_{ij}^2\mathbf{x}_i^T{\mathbf{\Theta} ^{{{(j)}^T}}}\mathbf{U}}  + \beta \sum\limits_{r\ne i} {{s_{ir}}} {\mathbf{v}_{r,:}})_k}}}}{{{{({\mathbf{v}^{(t)}_{i,:}}{\mathbf{U}^T}\mathbf{U}\sum\limits_{j = 1}^m {p_{ij}^2}  + \beta \sum\limits_{r\ne i} {{s_{ir}}{\mathbf{v}^{(t)}_{i,:}}} )_k}}}}} .
\end{equation}
According to Lemma~\ref{lemma1}, ${\cal L}(\mathbf{v}_{i,:})$ is non-increasing with the above updating rule.
\end{proof}

\section{Experiments}
\label{section_exper}
In this section, experiments on synthetic and real-world datasets are conducted to demonstrate the effectiveness of FNMF. Throughout the experiments, the number of components $m$ is fixed as 3 empirically.

\subsection{Results on Synthetic Dataset}
To investigate the effectiveness of feature weighting, a synthetic dataset is constructed. The dataset contains 900 samples with seven dimensions from three classes. In the first two dimensions, the samples from each class obey a specific Gaussian distribution, as shown in Fig.~\ref{toy1} (a). The last five noisy dimensions are randomly generated in 0 and 3.

We employ NMF~\cite{nmf}, RNMF~\cite{l21nmf} and GNMF~\cite{gnmf} as comparison methods, which will be described in detail in Section~\ref{section_rw}. After obtaining the new representation, $K$-means is employed to get the clustering results, as shown in Fig.~\ref{toy1} (b)-(e). NMF, RNMF and GNMF obtain incorrect clustering results because the noisy dimensions affect the representation learning. RNMF is proposed to deal with the outliers, but it fails when all the samples are with noisy features. GNMF relies on the data graph, which is unreliable when the noise is large. FNMF exploits the importances of features with the feature weighting components, so it is able to alleviate the noisy dimensions and achieve better clustering performance, as shown in Fig.~\ref{toy1} (e).

\begin{table}
\caption{Details of the real-world datasets.}
\label{tabledata}
\centering
\small
\renewcommand\arraystretch{1.2}
\begin{tabular}{|p{1.6cm}<{\centering}||p{1.4cm}<{\centering}|p{1.4cm}<{\centering}|p{1.4cm}<{\centering}|}

\hline
Datasets&  Classes &Samples &Features\\
\hline
\hline
YALE &15&165&256\\ \hline
ORL &40 &400 &1024\\ \hline
BA &36 &1404 &320\\  \hline
USPS &10& 1854& 256\\ \hline
Semeion &10 &1593& 256\\ \hline
CNAE-9 &9 &1080& 856\\ \hline
Glass &6 &214 &9\\ \hline
Mfeat &10 &2000& 240\\ \hline
\end{tabular}

\end{table}
\subsection{Results on Real-World Benchmarks}
\label{section_rw}
In this part, performance on real-world datasets is provided. We use the Clustering accuracy (ACC) and Normalized Mutual Information (NMI) as measurements.

\begin{table*}
\caption{ACC on real-world datasets. Best results are shown in bold face.}
\label{tableacc}
\centering
\small
\renewcommand\arraystretch{1.2}
\begin{tabular}{|p{1.6cm}<{\centering}||p{1.4cm}<{\centering}|p{1.4cm}<{\centering}|p{1.4cm}<{\centering}|p{1.4cm}<{\centering}|p{1.4cm}<{\centering}|p{1.4cm}<{\centering}|p{1.4cm}<{\centering}|p{1.4cm}<{\centering}|p{1.4cm}<{\centering}|}

\hline
Methods&  YALE &ORL & BA  & USPS & Semeion & CNAE-9 & Glass &Mfeat\\
\hline
\hline
NMF &0.4461&0.5670  &0.3151 &0.6046 &0.5188 &0.5818 &0.4519 &0.6416\\ \hline
RNMF  &0.4242&0.5505 &0.3450 &0.6820 &0.5255 &0.5867 &0.5238 &0.7087\\ \hline
GNMF &0.4473&0.6210  &0.4321 &0.7681 &0.6599 &0.5759 &0.4944 &0.9105\\ \hline
RMNMF  &0.4194&0.5270 &0.2887 &0.6098 &0.4328 &0.4957 &0.3991 &0.5456\\ \hline
LCNMF  &0.2303&0.4080 &0.0640 &0.5472 &0.2997 &0.2352 &0.4668 &0.6706\\ \hline
ONGR &0.4642&0.5350  &0.3929 &0.6841 &0.5675 &0.6108 &0.4243 &0.7766\\ \hline
CAN  &0.4182 &0.5650&0.3298 &0.7697 &0.5819 &0.6583 &0.5140 &0.8290\\ \hline
PCAN  &0.4121&0.5450 &0.2792 &0.7260 &0.5844 &0.6722 &0.5187 &0.8180\\ \hline
CLR  &0.4485&0.5225 &0.2251 &0.6915 &0.4551 &0.3889 &0.4626 &0.8660\\ \hline
KMM  &0.3576 &0.5220&0.2902 &0.7225 &0.5371 &0.5336 &0.5112 &0.8405\\ \hline
FNMF  &\textbf{0.4994}&\textbf{0.6350} &\textbf{0.4791} &\textbf{0.8302} &\textbf{0.6925} &\textbf{0.6815} &\textbf{0.5374} &\textbf{0.9219}\\ \hline
\end{tabular}

\end{table*}

\textbf{Datasets}: eight datasets are employed to evaluate the performance.
\begin{itemize}
\item[1)] \textbf{YALE}~\cite{yale} contains 165 face images captured from 15 persons. For each person, there are 11 images taken under different conditions, such as happy, normal, surprised, etc.

\item[2)] \textbf{ORL}~\cite{orl} is consisted of 400 face images of 40 persons. The images are with different lighting and expressions and facial details.

\item[3)] \textbf{Binary Alphadigits} (BA)\footnote{https://cs.nyu.edu/~roweis/data.html \label{roweis}} contains the binary digits of 0 to 9 and capitals A to Z. There are 36 classes in total, and each class has 39 images.

\item[4)] \textbf{USPS}~\cite{gnmf} is a widely used handwritten digits dataset. It contains gray scale images from 10 classes. We use a subset of USPS that contains 1854 samples.

\item[5)] \textbf{Semeion}~\cite{uci} has 1593 digits of 0 to 9 written by different persons. Each pixel was scaled to the 0/1 value with a threshold.

\item[6)] \textbf{CNAE-9}~\cite{uci} contains 1080 documents of business descriptions. Each document is represented as a vector, and the features are highly sparse with many zero values.

\item[7)] \textbf{Glass}~\cite{uci} records the oxide contents of 6 categories of glasses. There are 9 attributes for each sample corresponding to the contents of Na, Fe, etc.

\item[8)] \textbf{Multiple features} (Mfeat)~\cite{uci} collects the features of handwritten numerals. The features include Fourier coefficients, profile correlations, Zernike moments, etc.
\end{itemize}
In the experiments, each $\mathbf{x}_i$ is normalized as a unit vector. The datasets are described in detail in Table~\ref{tabledata}.

\textbf{Competitors}: ten data representation and clustering methods are used for comparison, which are listed as follows.
\begin{itemize}
\item[1)] \textbf{NMF}~\cite{nmf} is the classical model with the least square loss.

\item[2)] \textbf{Robust NMF} (RNMF)~\cite{l21nmf} is the robust version of NMF with the $\ell _{2,1}$ norm.

\item[3)] \textbf{Graph-regularized NMF} (GNMF)~\cite{gnmf} combines NMF with the local manifold.

\item[4)] \textbf{Robust Manifold NMF} (RMNMF)~\cite{rmnmf} integrates the advantages of RNMF and GNMF.

\item[5)] \textbf{Local Centroids-structured NMF}~\cite{lcnmf} (LCNMF) uses local centroids to approximate the samples within each class.

\item[6)] \textbf{Orthogonal and Nonnegative Graph Reconstruction} (ONGR)~\cite{ongr} accomplishes matrix factorization on the data graph.

\item[7)] \textbf{Constrained Adaptive Neighbors} (CAN)~\cite{can} learns the neighboring relationship with the Laplacian constraints.

\item[8)] \textbf{Projected CAN} (PCAN)~\cite{can} performs CAN in the projected subspace.

\item[9)] \textbf{Constrained Laplacian Rank} (CLR)~\cite{can} searches the optimal data graph based on the predefined graph.
 
\item[10)] \textbf{$K$-Multiple Means} (KMM)~\cite{kmm} introduces local centroids into $K$-means.
\end{itemize}
For the matrix factorization methods, i.e. 1)-6), $K$-means is performed on learned data representation for clustering. To avoid the influence of initialization, both the representation learning and clustering procedures are repeated for twenty times, and the averaged results are reported. For the clustering methods, i.e. 7)-10), the clustering result are obtained without post-processing. The parameters of all the algorithms, including the proposed FNMF, are selected by searching the grid $\{10^{-3},10^{-2},\cdots,10^3\}$. For FNMF, RMNMF, ONGR, CLR and FNMF, the graph is constructed with the algorithm in~\cite{clr} and the neighborhood size is set as 5.

\begin{figure}
\begin{center}
\subfigure[Features on YALE]{
\includegraphics[width=0.23\textwidth]{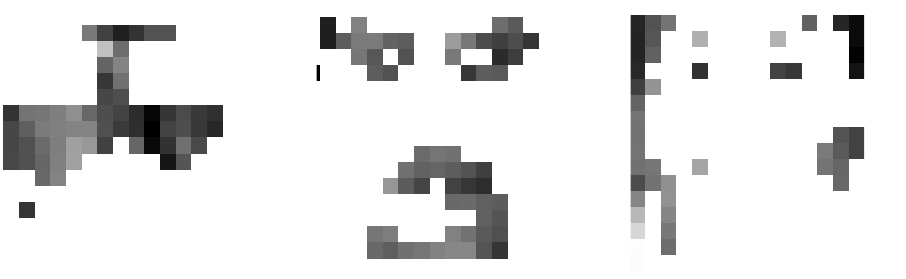}}
\subfigure[Features on ORL]{
\includegraphics[width=0.23\textwidth]{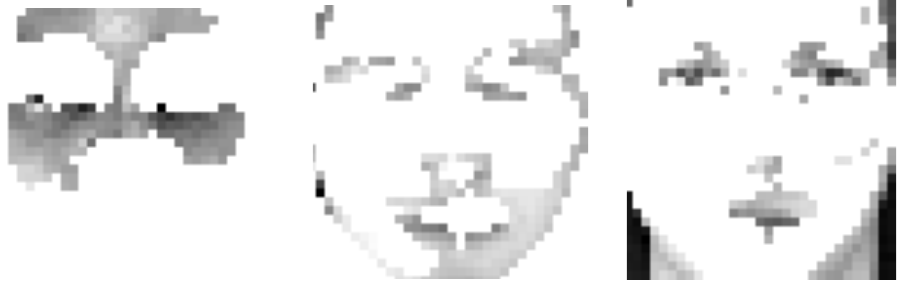}}
\end{center}
\caption{$25\%$ features selected by the components learned on  YALE and ORL.}
\label{facefeatures}
\end{figure}

\begin{figure}
\begin{center}
\subfigure[$4\times 4$]{
\includegraphics[width=0.46\textwidth]{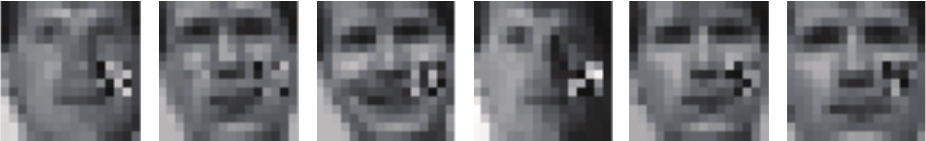}}
\subfigure[$6\times 6$]{
\includegraphics[width=0.46\textwidth]{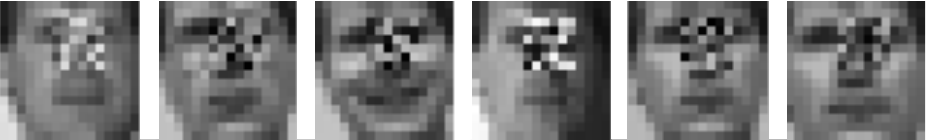}}
\subfigure[$8\times 8$]{
\includegraphics[width=0.46\textwidth]{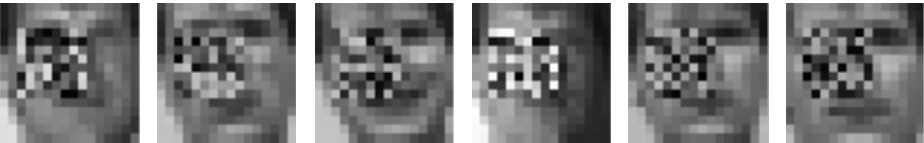}}
\end{center}
\caption{Samples with different size of occlusions from Yale dataset.}
\label{occs}
\end{figure}

\begin{table*}
\caption{NMI on real-world datasets. Best results are shown in bold face.}
\label{tablenmi}
\centering
\small
\renewcommand\arraystretch{1.2}
\begin{tabular}{|p{1.6cm}<{\centering}||p{1.4cm}<{\centering}|p{1.4cm}<{\centering}|p{1.4cm}<{\centering}|p{1.4cm}<{\centering}|p{1.4cm}<{\centering}|p{1.4cm}<{\centering}|p{1.4cm}<{\centering}|p{1.4cm}<{\centering}|p{1.4cm}<{\centering}|}

\hline
Methods & YALE& ORL & BA  & USPS & Semeion & CNAE-9 & Glass &Mfeat\\
\hline
\hline
NMF   &0.5040&0.7438 &0.4709 &0.5771 &0.4358 &0.5163 &0.2987 &0.6001\\ \hline
RNMF  &0.4816&0.7319 &0.5066 &0.6518 &0.4647 &0.5189 &0.3761 &0.6428\\ \hline
GNMF  &0.4834&0.7777 &0.5859 &0.7704 &0.6356 &0.507 &0.3588 &0.8779\\ \hline
RMNMF  &0.4668&0.6923 &0.4396 &0.5497 &0.3844 &0.4215 &0.2599 &0.5169\\ \hline
LCNMF  &0.2513&0.5484 &0.0555 &0.5231 &0.2317 &0.1561 &0.2225 &0.6862\\ \hline
ONGR &0.5159&0.7140  &0.5477 &0.751 &0.5875 &0.5687 &0.3312 &0.8463\\ \hline
CAN &0.4642&0.7101  &0.4463 &0.7669 &0.5924 &0.6421 &0.3849 &0.8796\\ \hline
PCAN  &0.4381&0.6923 &0.3485 &0.7348 &0.5526 &0.6322 &0.3199 &0.8339\\ \hline
CLR  &0.4548&0.6803 &0.2900 &0.7643 &0.4522 &0.3643 &0.3729 &0.8671\\ \hline
KMM  &0.4185 &0.6852&0.4188 &0.7550 &0.5187 &0.5039 &0.3769 &0.8412\\ \hline
FNMF  &\textbf{0.5524}&\textbf{0.7807} &\textbf{0.6332} &\textbf{0.8069} &\textbf{0.6532} &\textbf{0.6425} &\textbf{0.3828} &\textbf{0.8799}\\ \hline
\end{tabular}

\end{table*}

\begin{figure*}
\begin{center}
\subfigure[YALE]{
\includegraphics[width=0.23\textwidth]{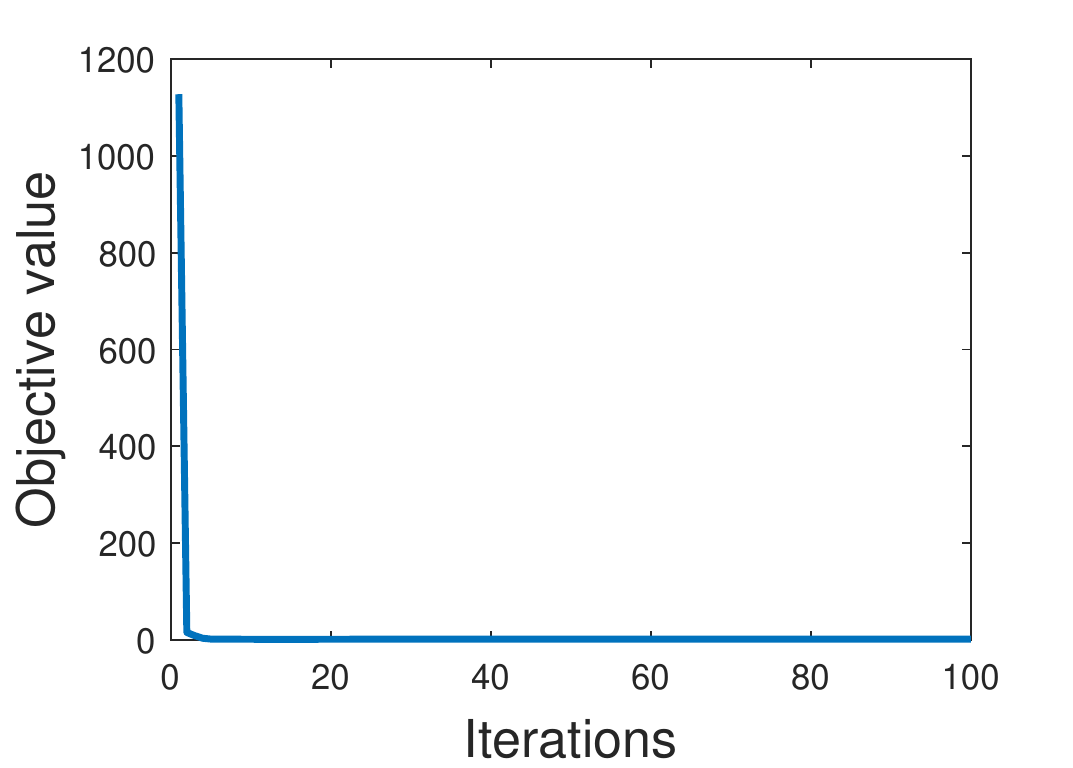}}
\subfigure[ORL]{
\includegraphics[width=0.23\textwidth]{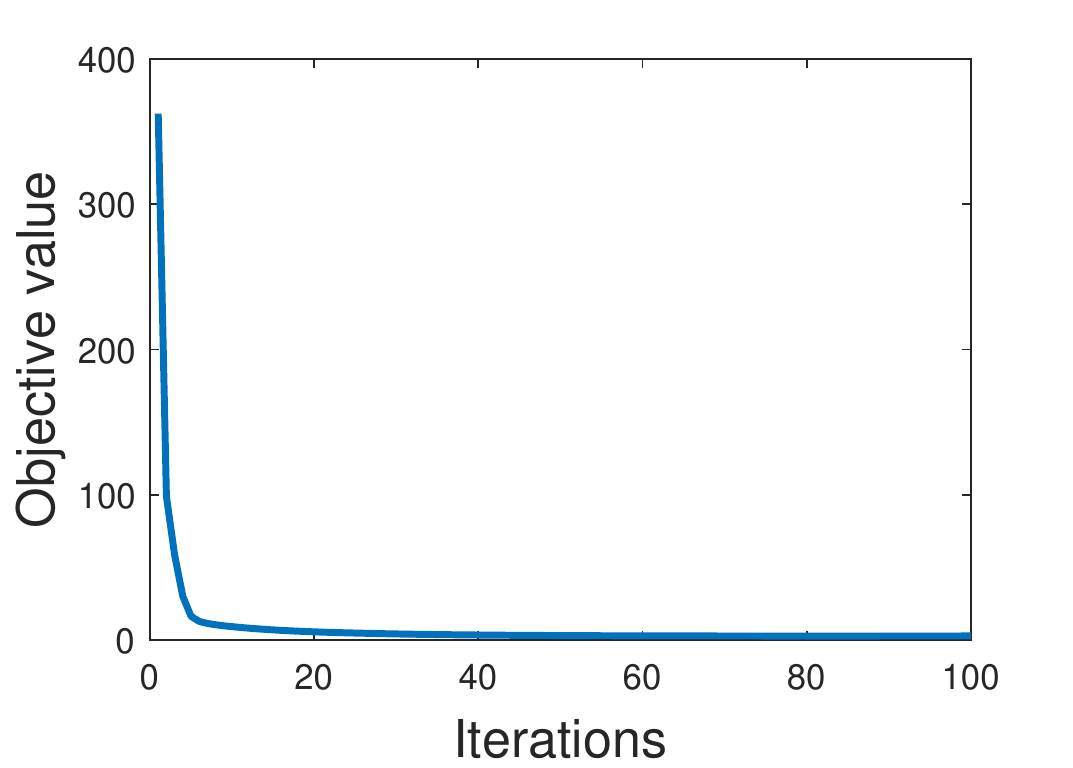}}
\subfigure[BA]{
\includegraphics[width=0.23\textwidth]{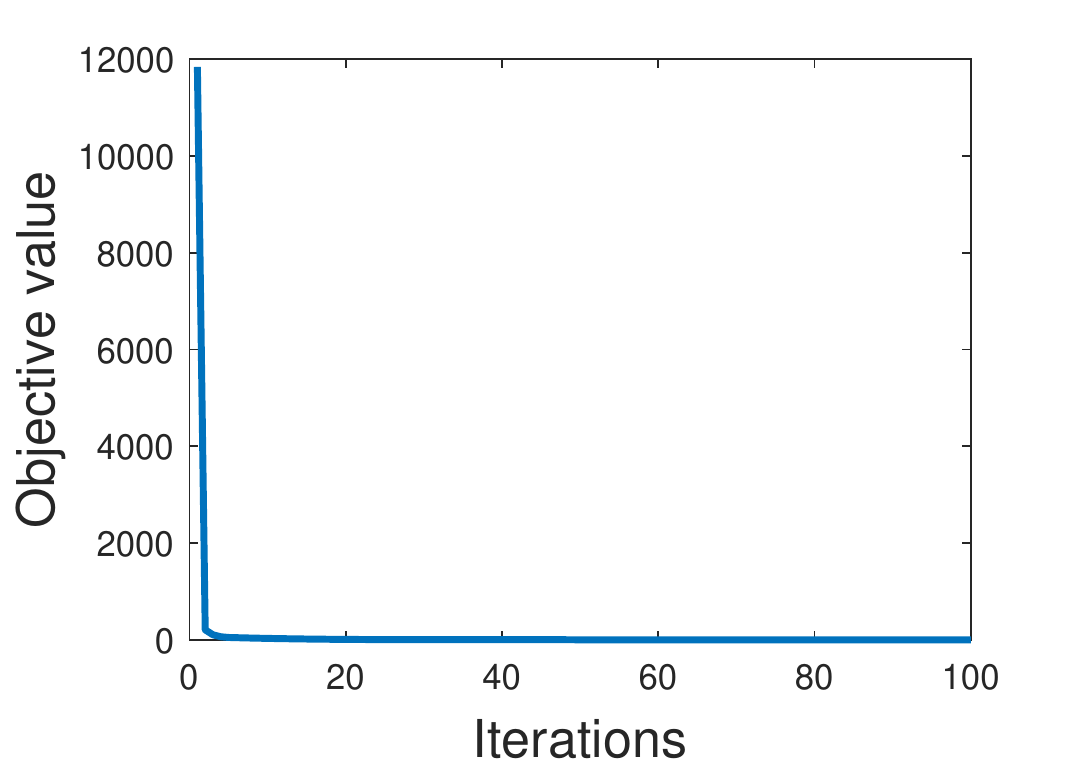}}
\subfigure[USPS]{
\includegraphics[width=0.23\textwidth]{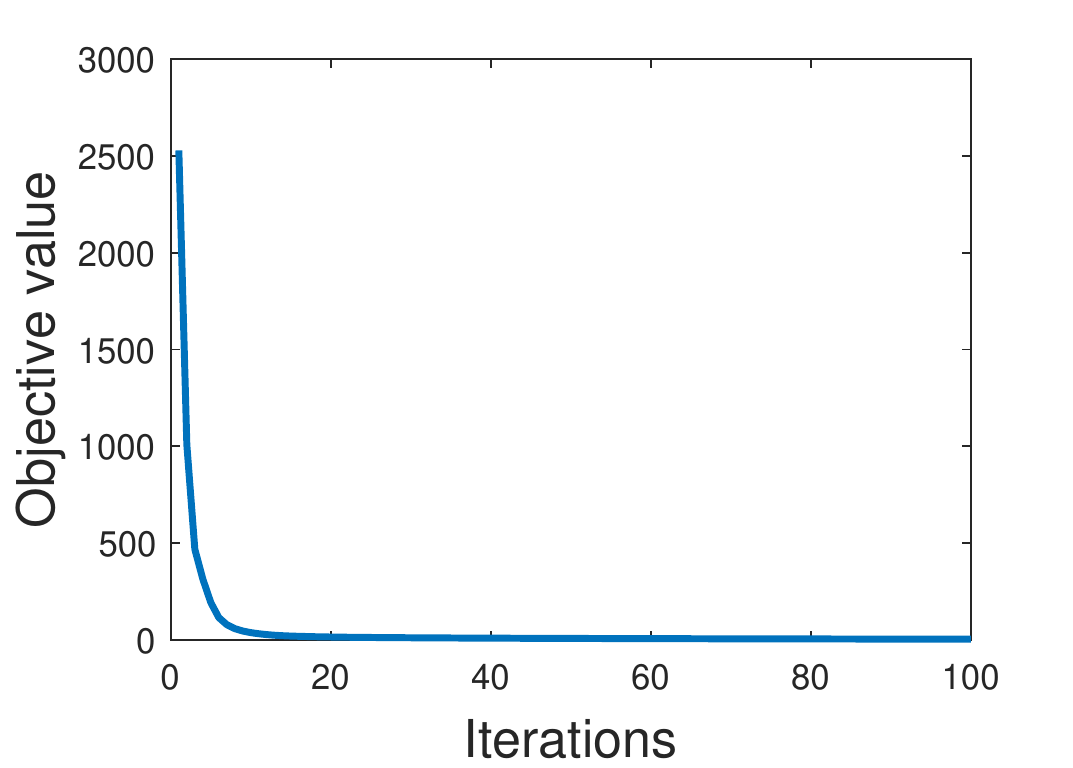}}
\end{center}

\begin{center}
\subfigure[Semeion]{
\includegraphics[width=0.23\textwidth]{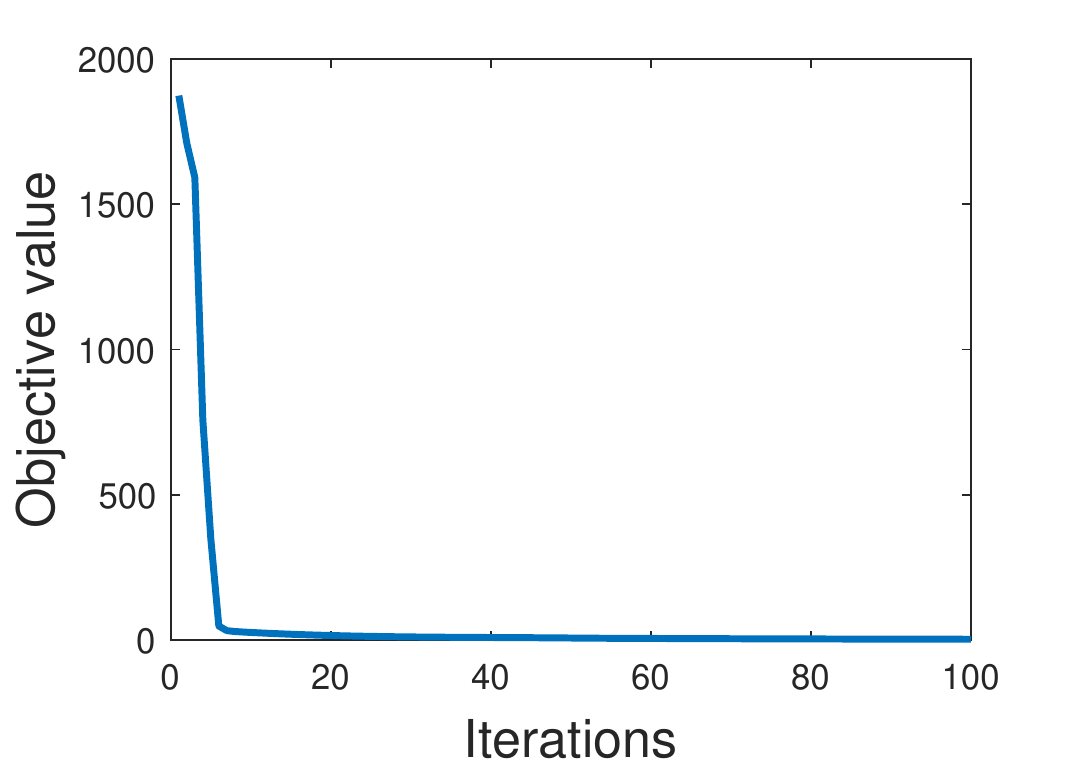}}
\subfigure[CNAE-9]{
\includegraphics[width=0.23\textwidth]{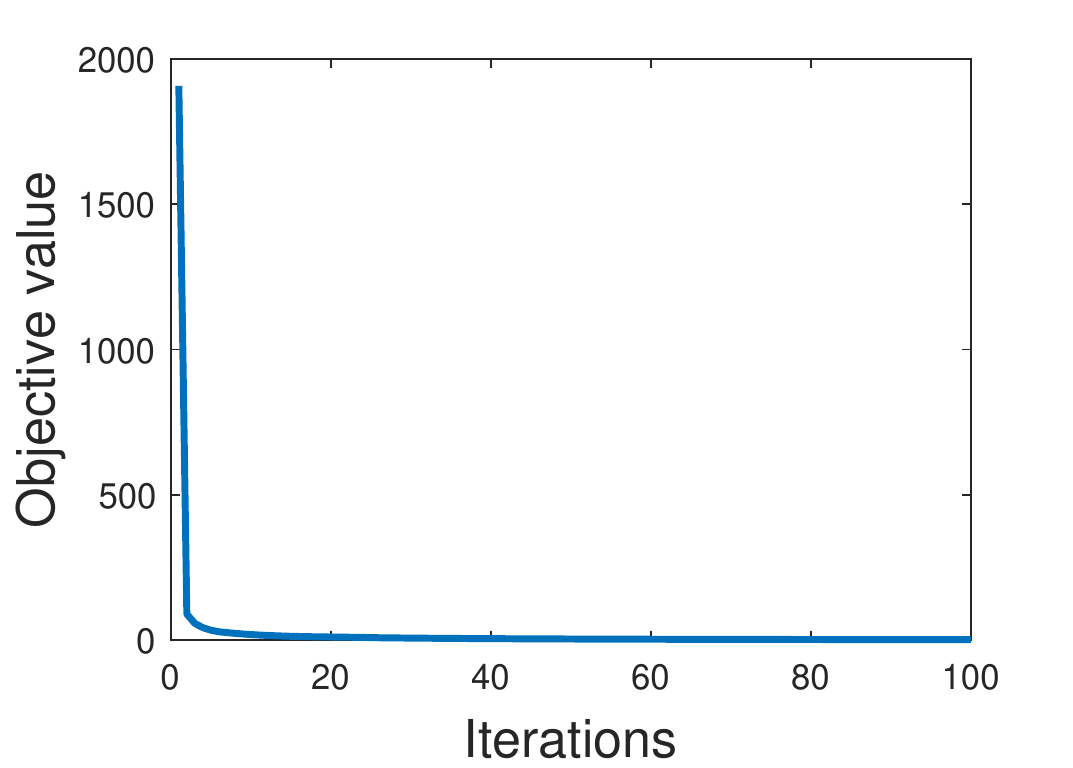}}
\subfigure[Glass]{
\includegraphics[width=0.23\textwidth]{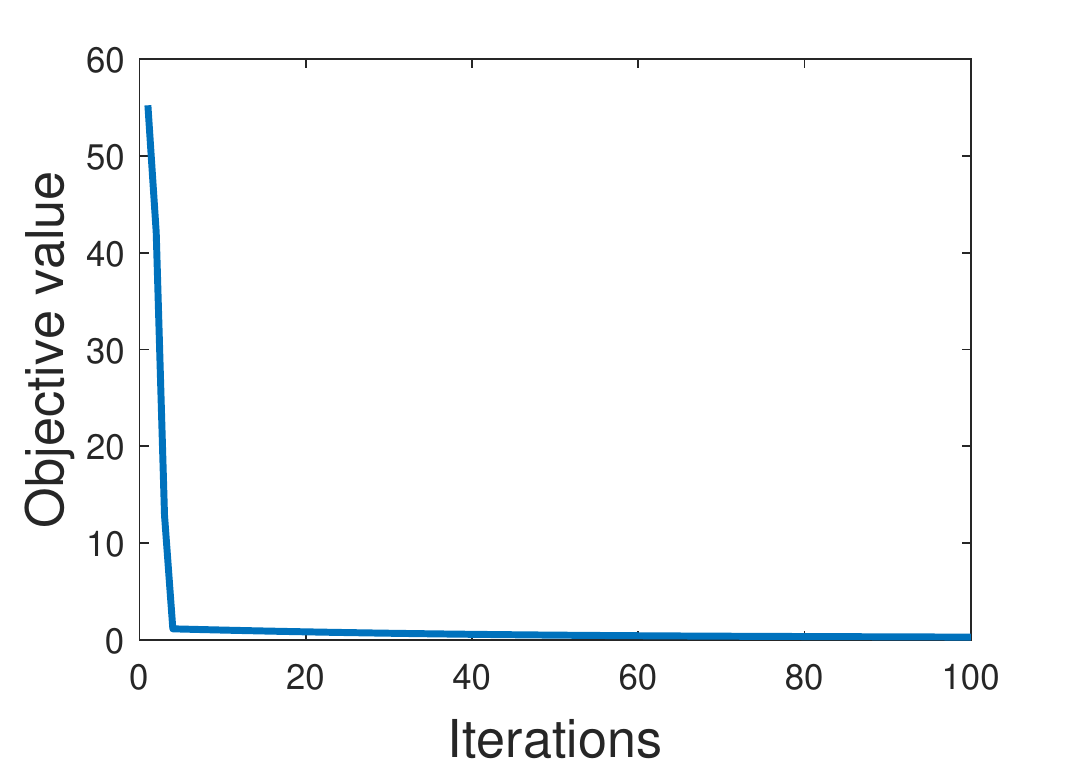}}
\subfigure[Mfeat]{
\includegraphics[width=0.23\textwidth]{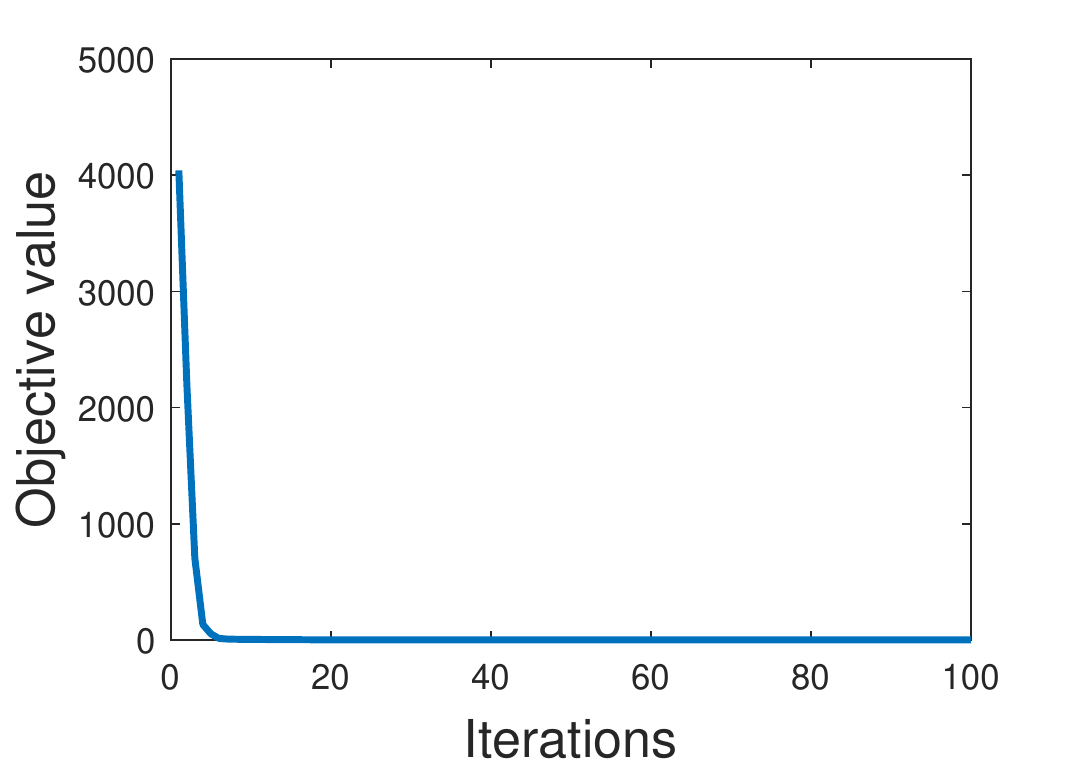}}
\end{center}
\caption{Convergence curves of FNMF on real-world datasets.}
\label{conv}
\end{figure*}

\begin{figure*}
\begin{center}
\subfigure[YALE]{
\includegraphics[width=0.23\textwidth]{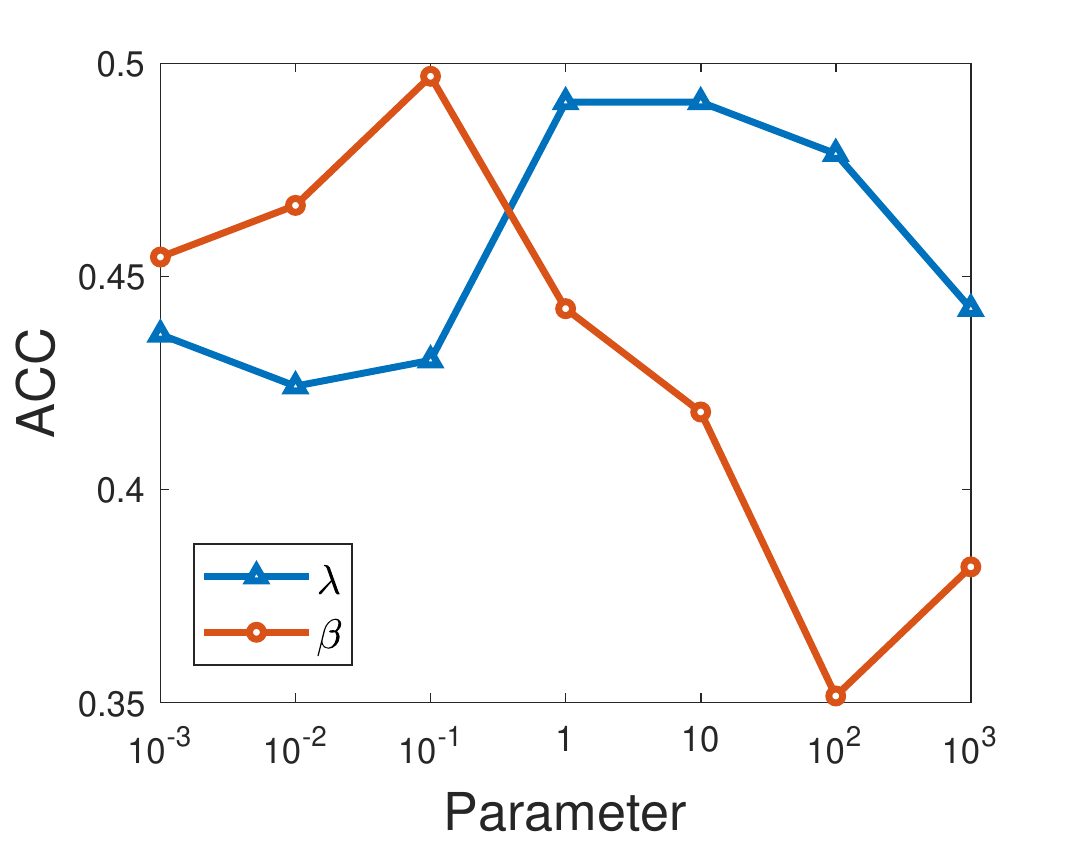}}
\subfigure[ORL]{
\includegraphics[width=0.23\textwidth]{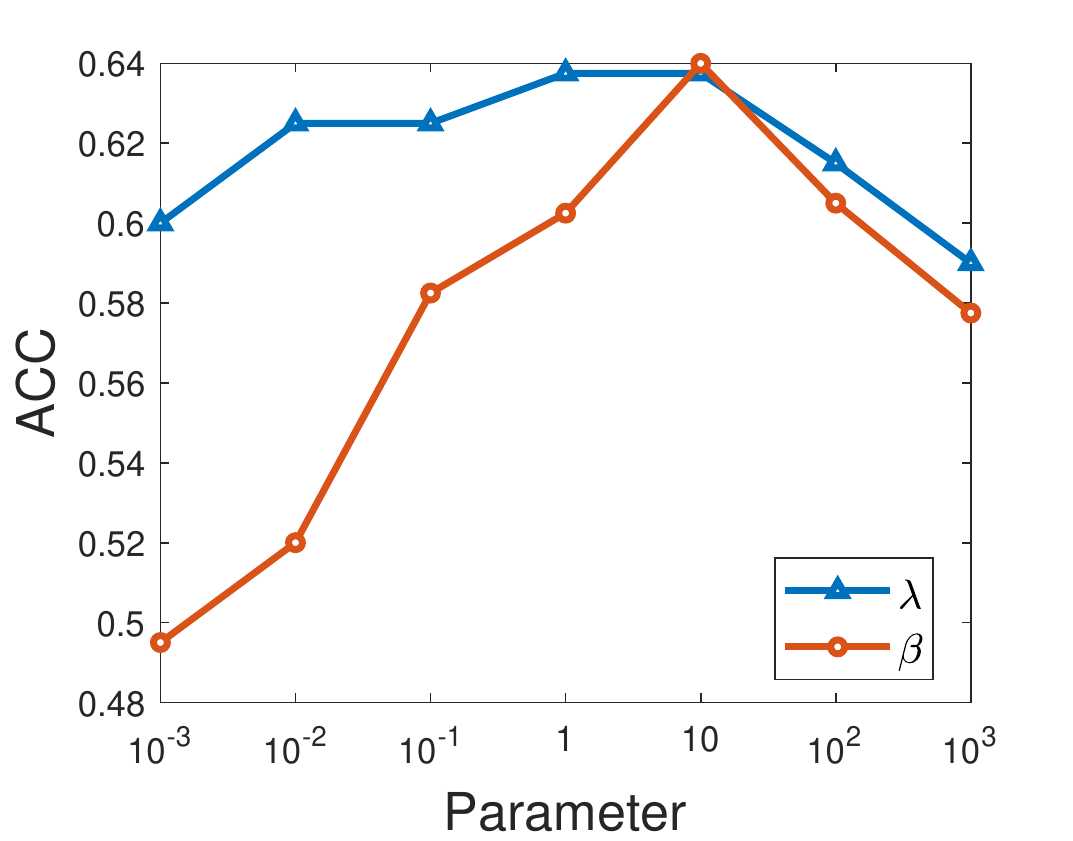}}
\subfigure[BA]{
\includegraphics[width=0.23\textwidth]{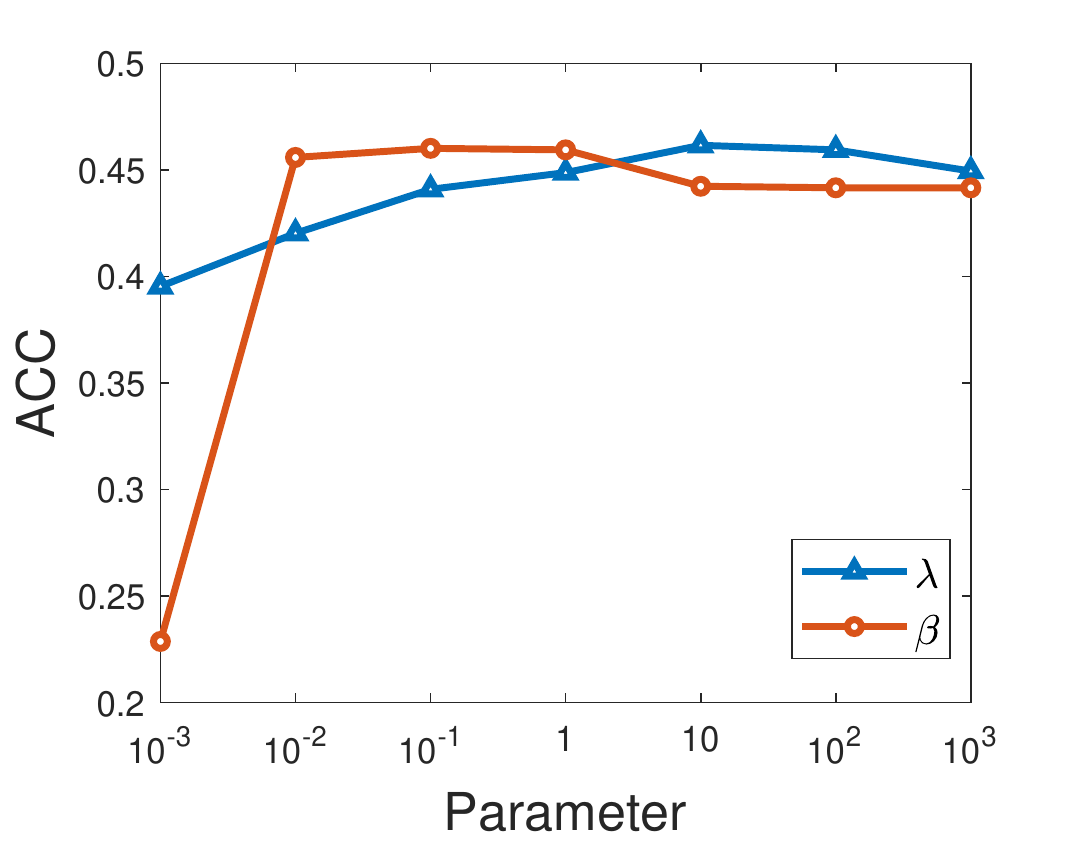}}
\subfigure[USPS]{
\includegraphics[width=0.23\textwidth]{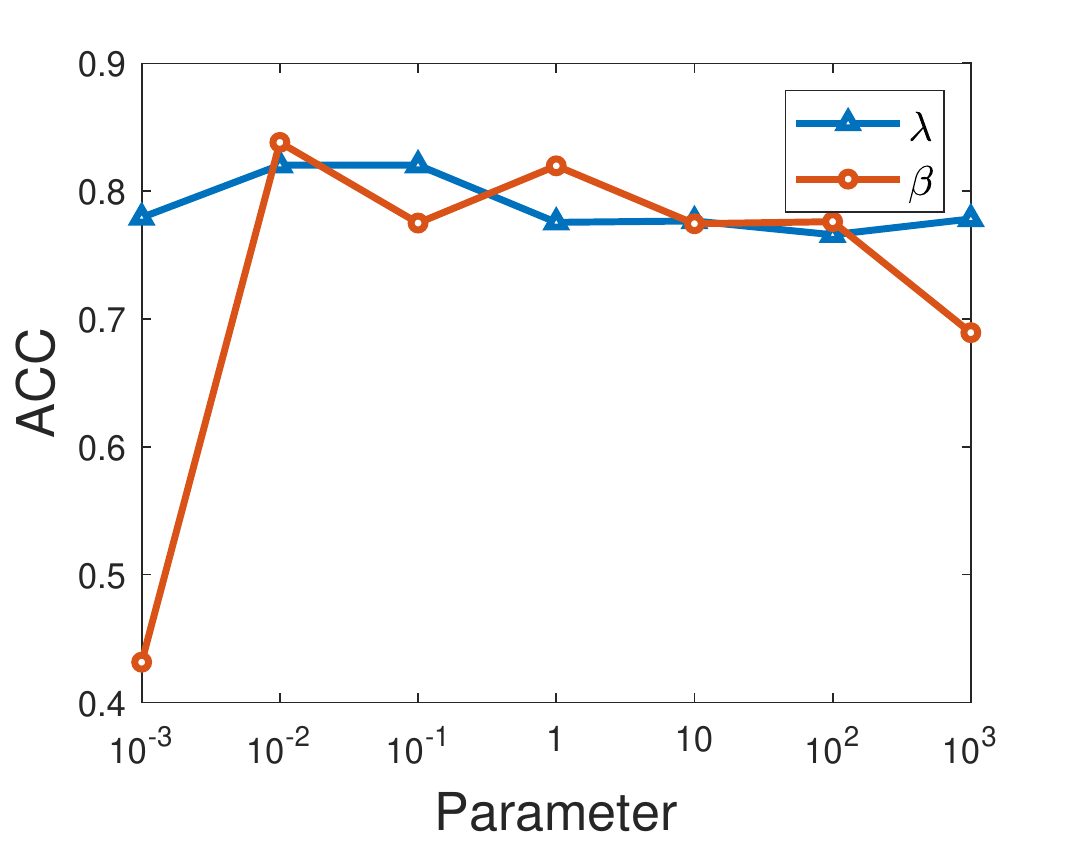}}
\end{center}

\begin{center}
\subfigure[Semeion]{
\includegraphics[width=0.23\textwidth]{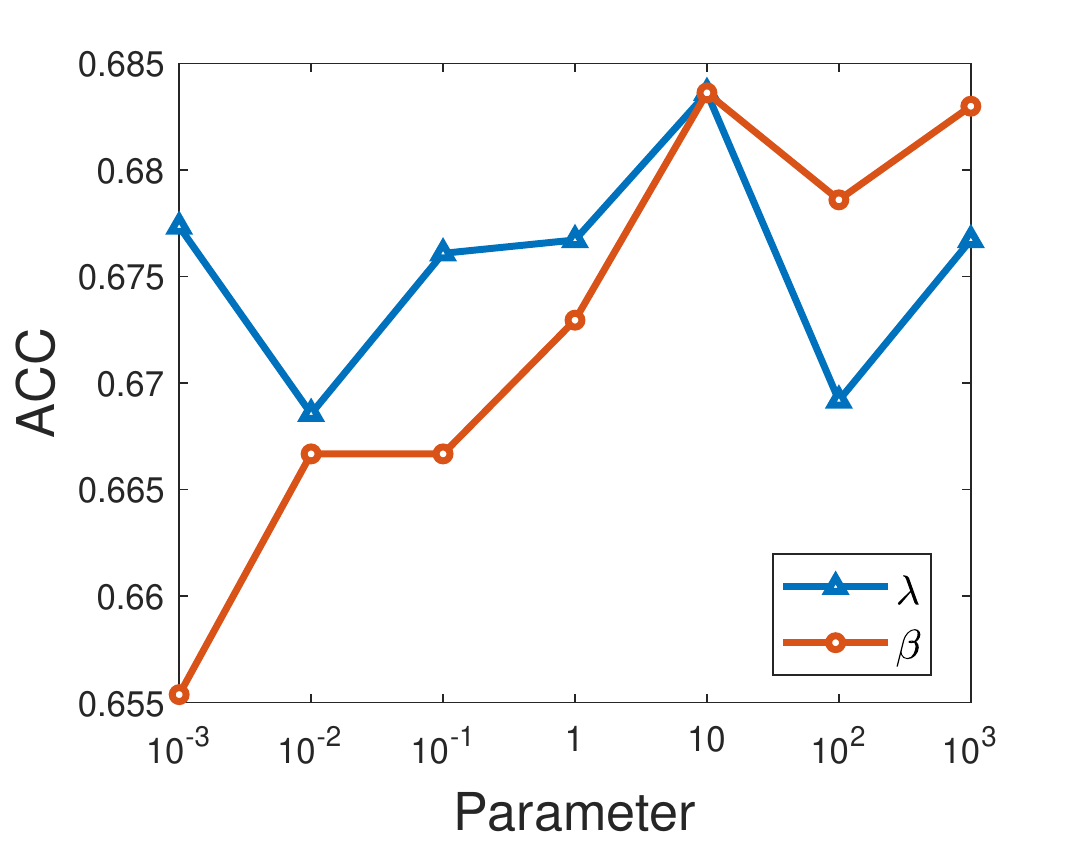}}
\subfigure[CNAE-9]{
\includegraphics[width=0.23\textwidth]{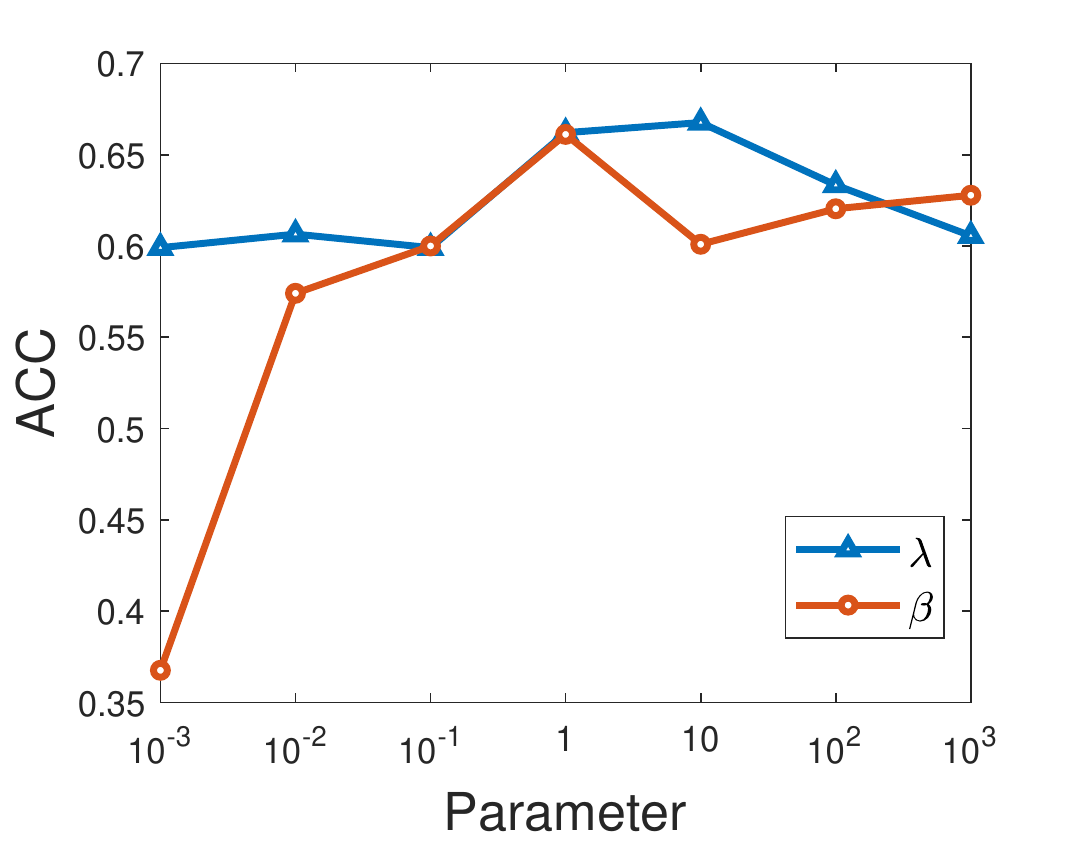}}
\subfigure[Glass]{
\includegraphics[width=0.23\textwidth]{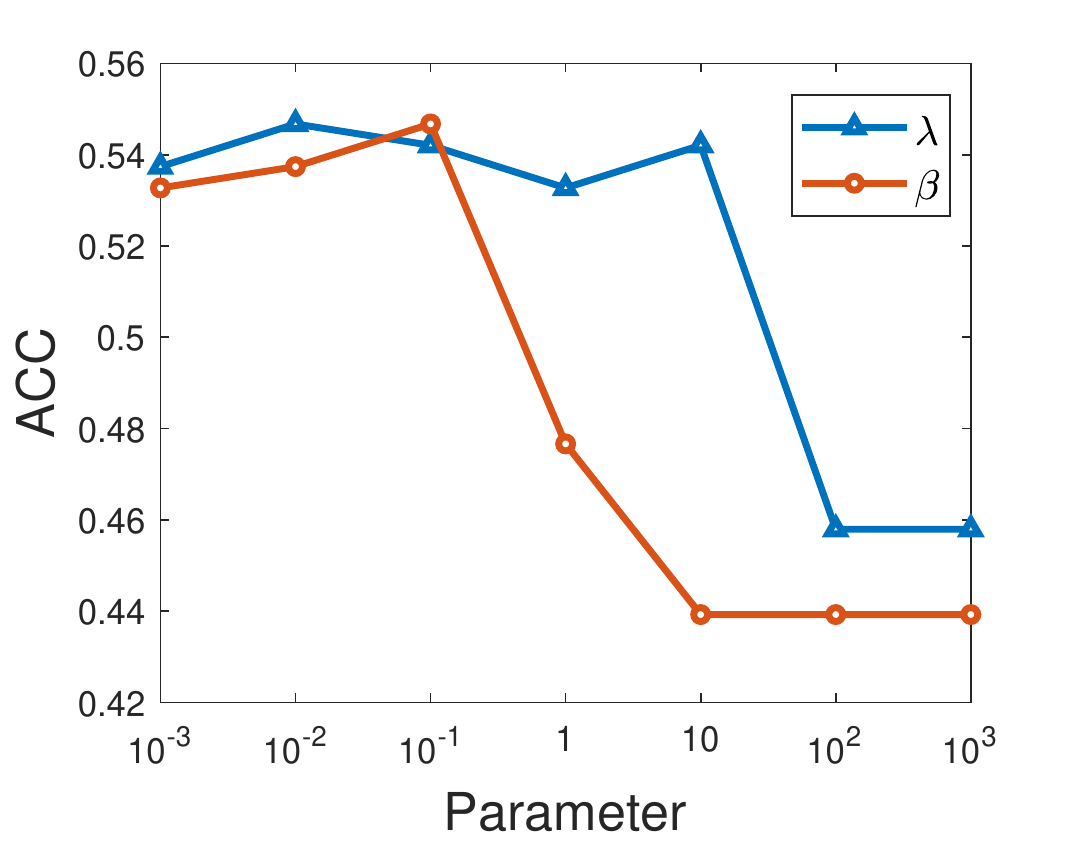}}
\subfigure[Mfeat]{
\includegraphics[width=0.23\textwidth]{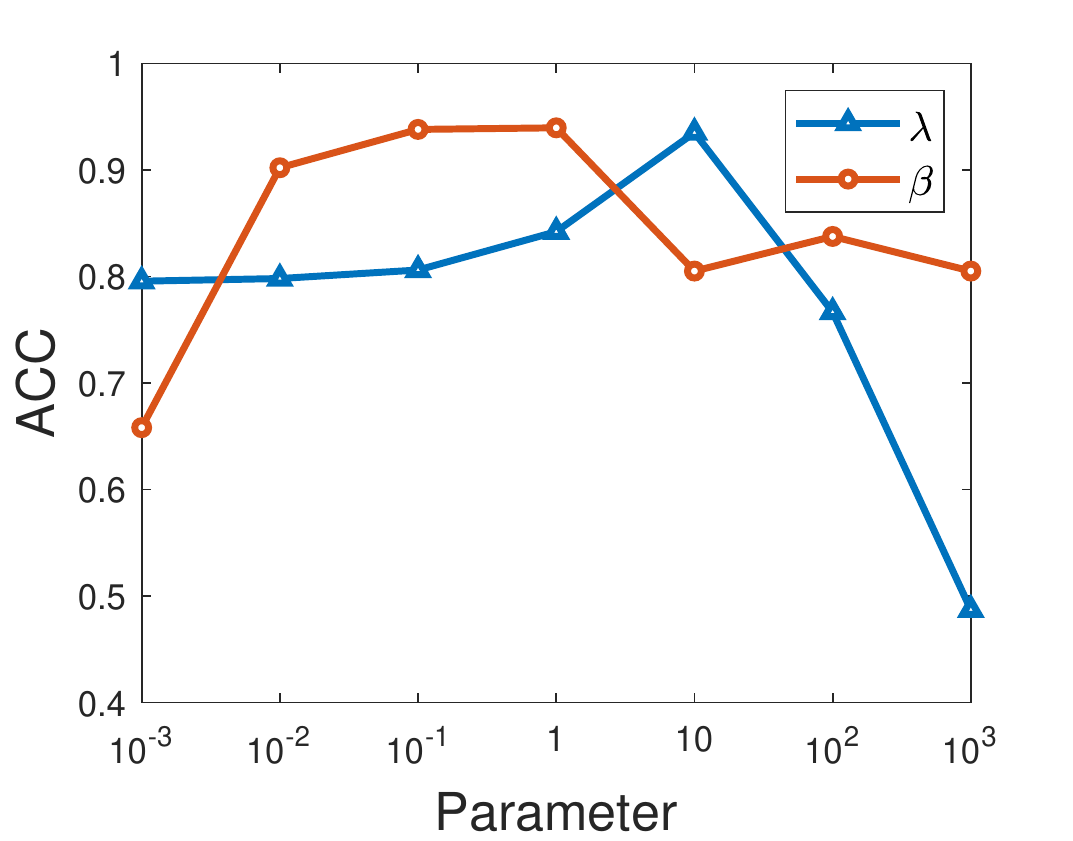}}
\end{center}
\caption{ACC of FNMF versus parameters $\lambda$ and $\beta$ on real-world datasets.}
\label{para}
\end{figure*}

\begin{figure*}
\begin{center}
\subfigure[YALE]{
\includegraphics[width=0.23\textwidth]{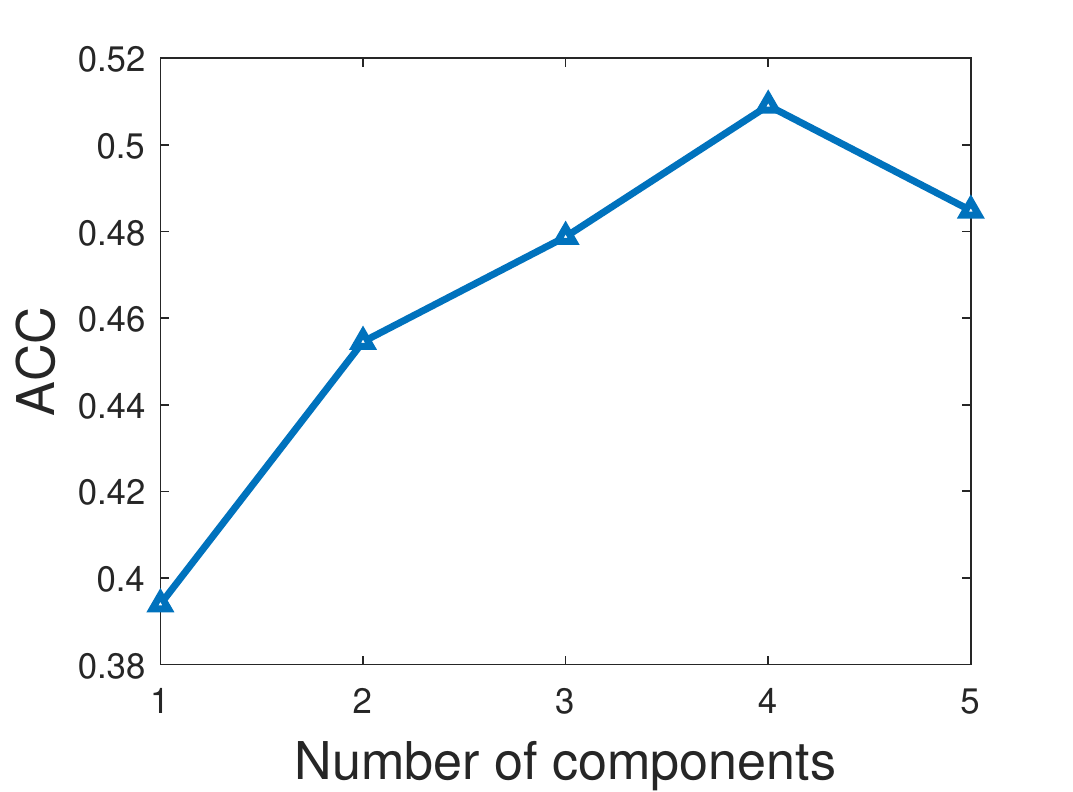}}
\subfigure[ORL]{
\includegraphics[width=0.23\textwidth]{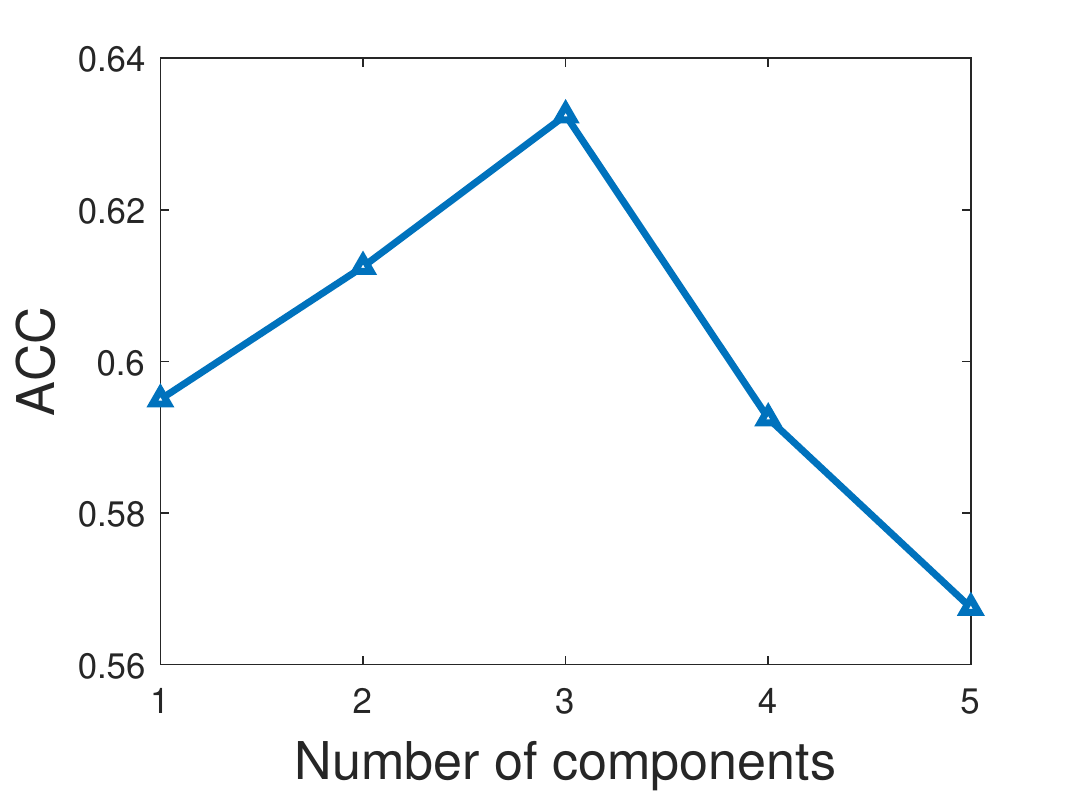}}
\subfigure[BA]{
\includegraphics[width=0.23\textwidth]{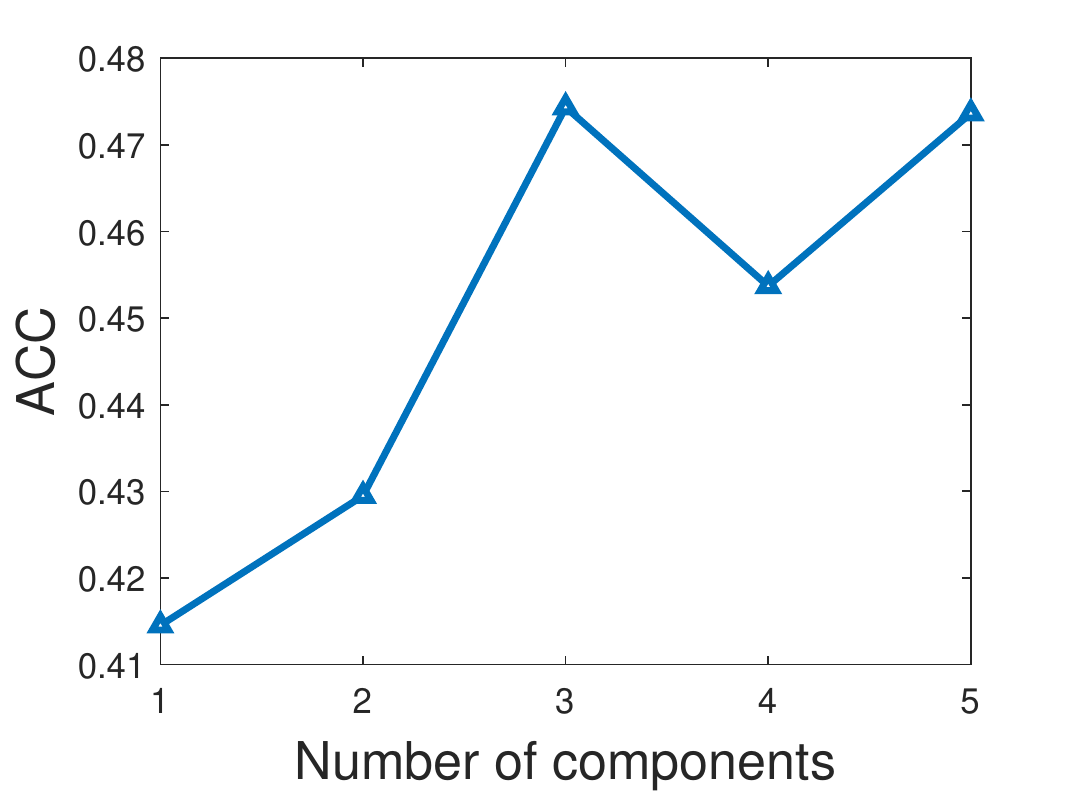}}
\subfigure[USPS]{
\includegraphics[width=0.23\textwidth]{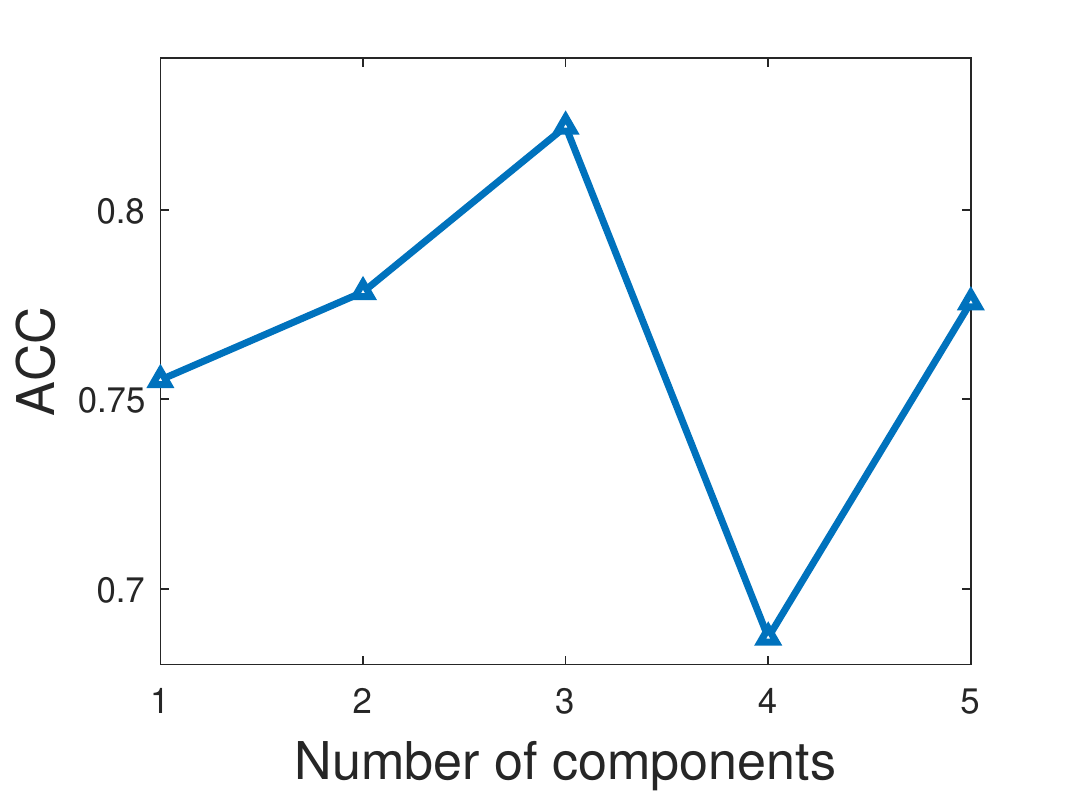}}
\end{center}

\begin{center}
\subfigure[Semeion]{
\includegraphics[width=0.23\textwidth]{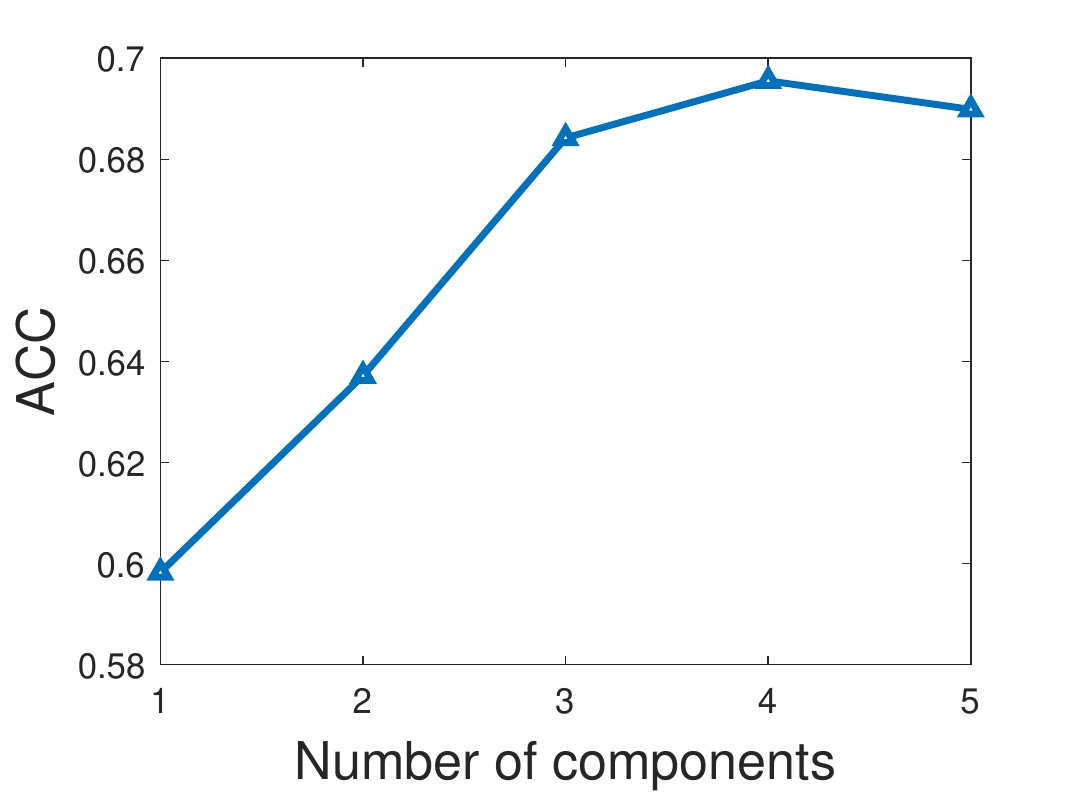}}
\subfigure[CNAE-9]{
\includegraphics[width=0.23\textwidth]{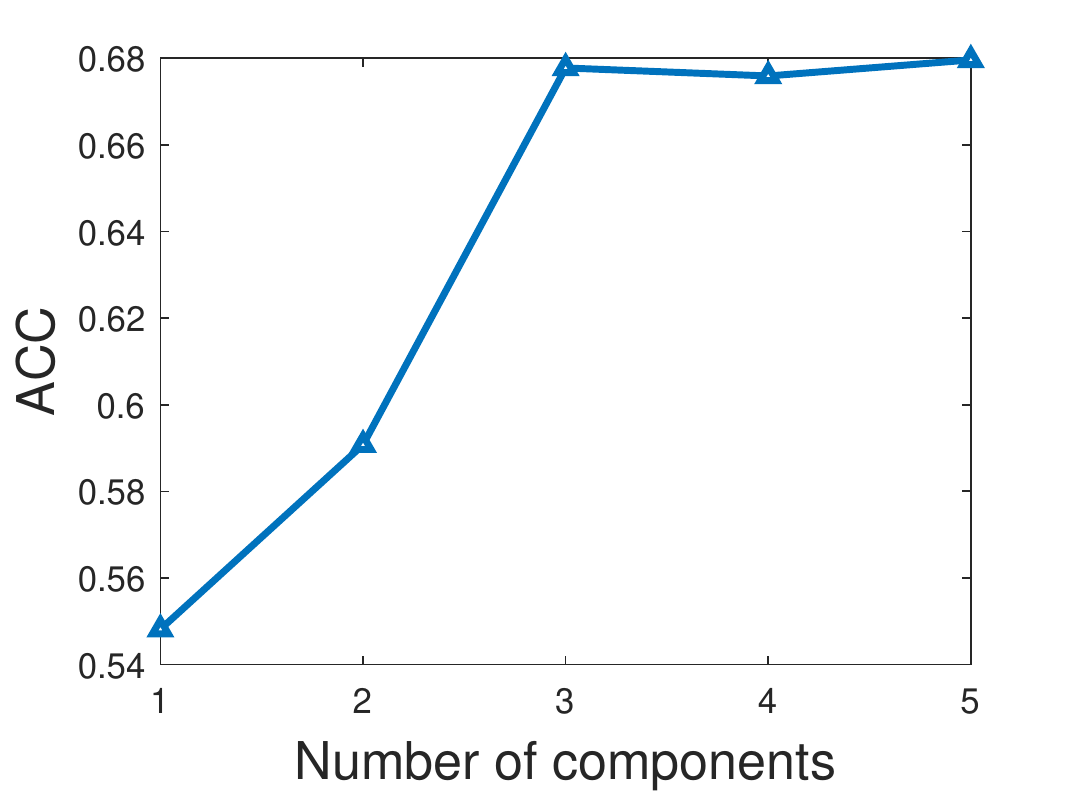}}
\subfigure[Glass]{
\includegraphics[width=0.23\textwidth]{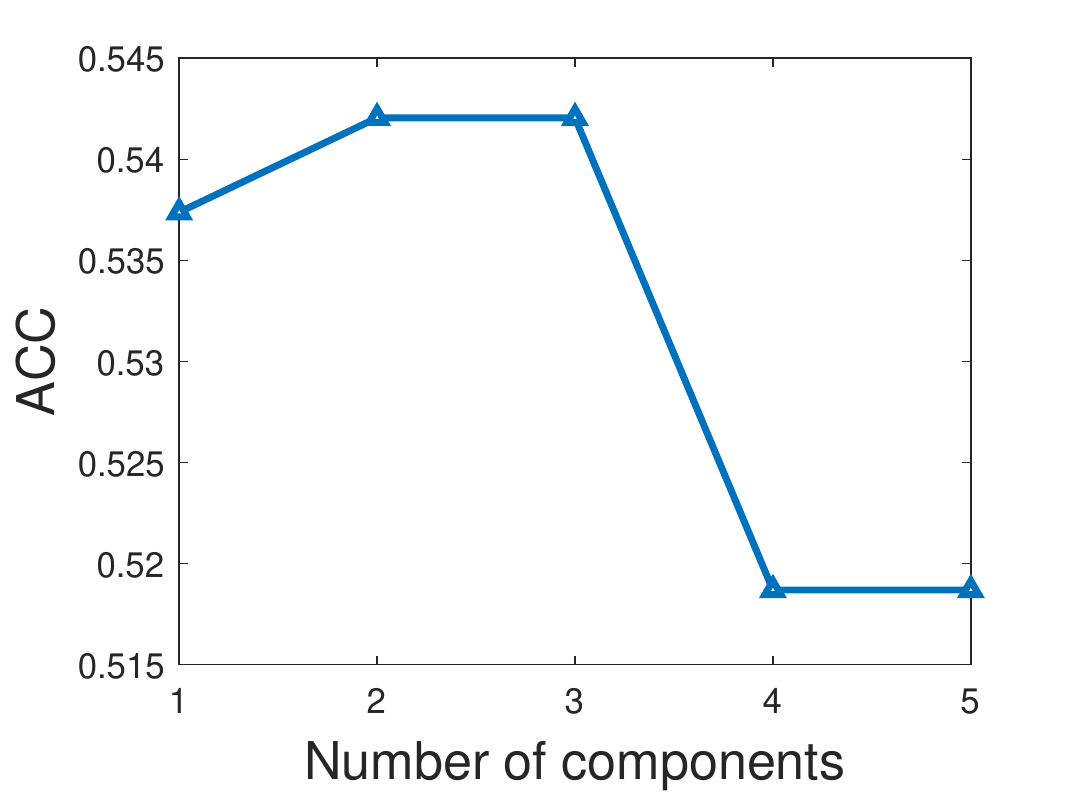}}
\subfigure[Mfeat]{
\includegraphics[width=0.23\textwidth]{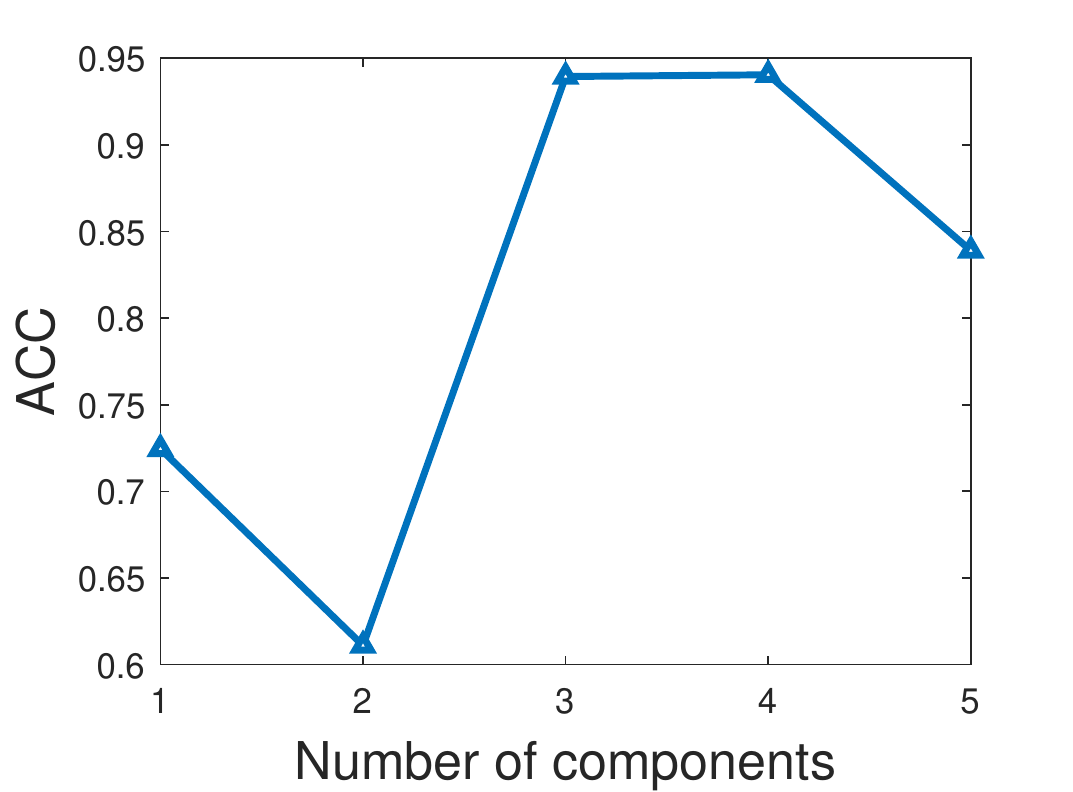}}
\end{center}
\caption{ACC of FNMF with different number of feature weighting components on real-world datasets.}
\label{features}
\end{figure*}

\textbf{Performance}: the results of different methods are exhibited in Table~\ref{tableacc} and~\ref{tablenmi}. FNMF achieves the best performance in terms of ACC and NMI. Generally, the variants of NMF outperform NMF on most occasions. The results yield the following observations.
\begin{itemize}
\item GNMF, RMNMF and ONGR perform better than NMF and RNMF, which verifies the necessity of capturing data relationship. 

\item Both LCNMF and KMM find local centroids to represent the input data. KMM outperforms LCNMF because the centroids are constrained to be close to the samples.

\item On CNAE-9, the data contains many irrelevant features, so PCAN shows good performance. But it is inferior to FNMF since it changes the original data characteristics.

\item Compared with the competitors, FNMF utilizes feature weighting components to rank the importance of original features. Therefore, it learns the informative representation and shows the best performance.
\end{itemize}
Fig.~\ref{facefeatures} visualizes the top $25\%$ features ranked by each component of FNMF. The components capture complementary discriminative features, such as eye, eye socket, shape of face and mouth. 

\begin{table}
\caption{ACC on the noised Yale datasets. Best results are shown in bold face.}
\label{tablenoise}
\centering
\small
\renewcommand\arraystretch{1.2}
\begin{tabular}{|p{1.3cm}<{\centering}||p{1.2cm}<{\centering}|p{1.2cm}<{\centering}|p{1.2cm}<{\centering}|p{1.2cm}<{\centering}|}

\hline
Methods& Noise & 4$\times$4& 6$\times$6& 8$\times$8\\
\hline
\hline
NMF  &0.3352 &0.3818 &0.3697 &0.3067\\ \hline
RNMF &0.2848 &0.3152 &0.2885 &0.2667\\ \hline
GNMF &0.2994 &0.3406 &0.3212 &0.2842\\ \hline
RMNMF &0.2691 &0.3091 &0.2903 &0.2624\\ \hline
LCNMF &0.1121 &0.1576 &0.1412 &0.1103\\ \hline
ONGR &0.2267 &0.3285 &0.2691 &0.2297\\ \hline
CAN &0.2182 &0.2788 &0.2667 &0.2364\\ \hline
PCAN &0.1939 &0.2485 &0.2364 &0.2000\\ \hline
CLR &0.2424 &0.3091 &0.2727 &0.2182\\ \hline
KMM &0.2158 &0.2836 &0.2206 &0.1933\\ \hline
FNMF &\textbf{0.3903} &\textbf{0.4406} &\textbf{0.4067} &\textbf{0.3964}\\ \hline
\end{tabular}

\end{table}

To further demonstrate the advantage of feature weighting, we add four kinds of noise features to the Yale datasets. For the first category, $\frac{d}{3}$ noisy dimensions are directly concentrated into the data matrix $\mathbf{X}$. For the last three categories, block noise with different sizes are added into the images, as shown in Fig.~\ref{occs}. All the noise features are randomly generated from 0 to the maximum value in $\mathbf{X}$. The results on the noised Yale datasets are given in Table~\ref{tablenoise}. FNMF outperforms the comparison methods in all cases. Therefore, FNMF is able to remove the noisy features while preserving the informative ones.

The convergence curves on the real-world datasets are shown in Fig.~\ref{conv}. The objective value converges within twenty iterations on all the datasets. The ACC curves with different value of $\lambda$ and $\beta$ are plotted in Fig.~\ref{para}. The results are stable across a wide range of parameter values. When $\lambda$ and $\beta$ are very large, the performance tend to decreases because the matrix approximation error increases.

In addition, we also investigate the necessity of introducing multiple feature weighting components. As shown in Fig.~\ref{features}, the performance is improved when $m$ increases from 1 to 3, which means that one single component is insufficient to reveal the importances of features. When $m$ exceeds 3, the performance on ORL and Glass decreases since some discriminative features are disassembled.

\section{Conclusions}
\label{section_conclusions}
In this paper, we put forward a new Feature-weighted Non-negative Matrix Factorization (FNMF) approach. Different from the existing approaches, FNMF learns the data representation with the auto-weighted features. The importance of the features are learned adaptively, and the diversity is preserved with multiple feature weighting components. By performing feature weighting and matrix factorization simultaneously, FNMF is able to select the most informative features when producing new representation. Extensive experiments on various datasets validate the superiority of FNMF, and show its capability on capturing discriminative features.

In the future work, we plan to develop the deep model of FNMF, such that it can be applied into large-scale datasets. Besides, it is also desirable to extend FNMF to the semi-supervised learning scheme.

%\section*{Acknowledgments}
%This work was supported by The National Key Research and Development Program of China under Grant 2018YFB1107400, and The National Natural Science Foundation of China under U1801262, 61871470, U1864204 and 61773316.

\bibliographystyle{IEEEtran}
\bibliography{IEEEabrv,Reference}

\end{document}